%% file: main.tex
\DocumentMetadata{testphase=new-or-1}
\documentclass[11pt]{article}
\usepackage[table]{xcolor}
\usepackage{tikz}
\usepackage{enumitem}
\usepackage{setspace}
\usepackage{booktabs}
\usepackage[margin=1in]{geometry}
\usepackage{amsmath}
\usepackage{csquotes}
\usepackage{quotes}
\usepackage{graphicx}
\usepackage[colorlinks=false, linkcolor=blue]{hyperref}
\usepackage[authordate,natbib=true,doi=false,isbn=false,url=false,uniquename=false]{biblatex-chicago}
\usetikzlibrary{shapes.geometric, arrows.meta, positioning, backgrounds, fit}
\usepackage{microtype}
\usepackage{subcaption}
\usepackage[nolists, nomarkers]{endfloat}
\usepackage{etoolbox}
\usepackage[flushleft]{threeparttable}
\AtBeginEnvironment{quote}{\par\singlespacing\small}
\newcommand{\grayrule}{\arrayrulecolor{black!20}\midrule\arrayrulecolor{black}}
\newcommand\fignote[1]{\captionsetup{font=footnotesize}\caption*{#1}}

\hypersetup{
     colorlinks   = true,
     citecolor    = blue,
     linkcolor     = blue
}

\DeclareFieldFormat{citehyperref}{%
  \DeclareFieldAlias{bibhyperref}{noformat}
  \bibhyperref{#1}}

\DeclareFieldFormat{textcitehyperref}{%
  \DeclareFieldAlias{bibhyperref}{noformat}
  \bibhyperref{%
    #1%
    \ifbool{cbx:parens}
      {\bibcloseparen\global\boolfalse{cbx:parens}}
      {}}}

\savebibmacro{cite}
\savebibmacro{textcite}

\renewbibmacro*{cite}{%
  \printtext[citehyperref]{%
    \restorebibmacro{cite}%
    \usebibmacro{cite}}}

\renewbibmacro*{textcite}{%
  \ifboolexpr{
    ( not test {\iffieldundef{prenote}} and
      test {\ifnumequal{\value{citecount}}{1}} )
    or
    ( not test {\iffieldundef{postnote}} and
      test {\ifnumequal{\value{citecount}}{\value{citetotal}}} )
  }
    {\DeclareFieldAlias{textcitehyperref}{noformat}}
    {}%
  \printtext[textcitehyperref]{%
    \restorebibmacro{textcite}%
    \usebibmacro{textcite}}}

\begin{document}
\title{Can LLMs Credibly Transform the Creation of Panel Data from Diverse Historical Tables?\thanks{ 
We thank Gordon Hanson, Jeffrey Lin, and Allison Shertzer for their helpful comments. Madison Dyhre Hansen, Nassir Holden, Svyatoslav Karnasevych, and Nathan Schor provided valuable research assistance.\\ 
\indent {\bf Disclaimer:} This paper represents research that is being circulated for discussion purposes. The views expressed here are solely those of the authors and do not necessarily reflect those of the Federal Reserve Bank of Philadelphia or the Federal Reserve System. Nothing in the text should be construed as an endorsement of any organization or its products or services. All errors or omissions are the responsibility of the authors. No statement here should be treated as legal advice.
}}

\author{Ver\'{o}nica B\"{a}cker-Peral\thanks{Massachusetts Institute of Technology: \href{mailto:vbperal@mit.edu}{vbperal@mit.edu}}
\and Vitaly Meursault\thanks{Federal Reserve Bank of Philadelphia: \href{mailto:vitaly.meursault@phil.frb.org}{vitaly.meursault@phil.frb.org} }
\and Christopher Severen\thanks{Federal Reserve Bank of Philadelphia: \href{mailto:chris.severen@phil.frb.org}{chris.severen@phil.frb.org} } }
\date{June 2026}

\maketitle

\begin{abstract}
\noindent 
Multimodal LLMs offer the potential for a watershed change for the digitization of historical tables by enabling low-cost processing that is centered on domain expertise rather than technical skill. We develop and rigorously assess an LLM-based pipeline on a new panel of historical county-level vehicle registration tables from early 20th-century U.S.\ state reports. Using human-transcribed gold standard data for evaluation, the pipeline achieves an exact cell match rate of 95.4\% at approximately 50 times less expense than traditional outsourcing. The pipeline performs well at extracting table structure, where it reduces critical parsing errors from 61.4\% to 0.35\%; in numerical transcription, where it exactly matches 96.7\% of linked cells and achieves a mean absolute percentage error of 0.7\%. The pipeline performs on par with human-based category alignment. We also assess pipeline performance in situ with two case studies that analyze the growth and persistence of historical vehicle adoption using common regression models. The significance and sign of effects are identical whether using LLM or gold standard data for all eight models tested, and the coefficient of interest is statistically indistinguishable in six of eight models. \\



\end{abstract}

\onehalfspacing

\newpage


\section{Introduction}

The advent of multimodal Large Language Models (LLMs) catalyzes a watershed moment in extracting historical information for quantitative analysis, particularly by facilitating extraction of complex, heterogeneous historical tables at scale.\footnote{We focus on machine-printed  historical tables. We do not address handwritten-text recognition (HTR) or severely degraded documents; the pipeline-building principles we describe apply in those settings as well, but the specific stage implementations and prompts (and hence our performance figures) may not.} The tables examined here---early 20th century automobile registration data produced by decentralized U.S.\ state agencies---exemplify heterogeneity through dramatic variation in layout structures, column naming conventions, geographic aggregations, and numeric representation conventions. Such tables lack the standardization common in other contexts (e.g., corporate balance sheets). While table digitization research has increasingly tackled diverse layouts and formats \citep[e.g.,][]{carlsonEfficientOCRBuilding2024, Correia2023, silcockNewswireLargeScaleStructured2024, Circi2024}, the salience of heterogeneity in historical administrative sources poses distinct challenges. Not only does successful analysis require accurate cell extraction from highly varied layouts, it also demands extensive cross-document harmonization to reconcile semantic differences and construct cohesive panels. 

We introduce and evaluate a digitization architecture that highlights the potential of multimodal LLMs for this task. This architecture is suited to transforming historical table scans into cohesive panel data. Unlike approaches that rely on customized machine learning or deep learning models and thus demand significant technical expertise in computer vision or natural language processing, this pipeline leverages the integrated visual and textual understanding of multimodal LLMs to handle complex tabular structures. Crucially, researchers guide the process using natural language prompts, applying their domain and data expertise (e.g., historical reporting conventions, geographical variations) to iteratively refine digitization and harmonization based on observed errors. Refinement driven by  domain knowledge proves particularly advantageous for settings where inconsistencies across source materials complicate both extraction and harmonization. Furthermore, the approach can offer substantial efficiency gains (estimated to be 1/50th the cost of outsourcing with less labor than manual processing), making complex panel dataset creation significantly more accessible to domain experts, regardless of technical background or budget.

To demonstrate the effectiveness of the LLM-driven pipeline, we apply it to a challenging test case: early 20th century county-level vehicle registration tables from the United States. These tables, produced independently by various state-level agencies, exhibit precisely the kind of heterogeneity that makes traditional digitization difficult, thus serving as a useful benchmark. Given the pipeline's sequential complexity (layout, extraction, harmonization) and the risk of compounding errors, rigorous end-to-end evaluation is essential. To enable pipeline development and evaluation, we manually created a ``gold standard'' dataset from 364 diverse (and often multi-page) tables encompassing 50,653 numerical cells. Using separate subsets of the data for prompt development and overall evaluation (79 and 285 tables), we comprehensively assess performance. Compared to standard OCR solutions often used as a baseline, the LLM pipeline substantially reduces critical layout parsing failures (that render tables unusable without substantial manipulation) from 61.4\% to 0.35\%. The pipeline achieves an exact cell match rate of 95.4\%. Among the linked cells, fidelity is high---the average error is very small (mean absolute percentage error 0.7\%) and 99.4\% of cells are within 10\% of the validated value. In 1.4\% of cases, the pipeline does not produce a value where the gold standard provides one; examining these cases reveals that they arise mostly from rare, idiosyncratic table formats and, to a much lesser extent, from header or county matching. The pipeline also performs on par with researchers at the subjective task of aligning field (header) labels, as validated against federally published state-total data.

Beyond cell-level performance, a critical question for applied research is whether any remaining errors propagate and bias downstream inference. We assess this by comparing standard econometric analyses using the LLM data and the gold standard data to study local vehicle adoption in the early 20th century, a topic of significant economic and historical importance \autocite{eli2022transportation}. Expansive sub-state level analysis has previously been unavailable due to data limitations, making these new insights particularly valuable.\footnote{A related innovation, the tractor, transformed agricultural labor \autocite{olmstead2001reshaping, manuelli2014frictionless}, and yet cars likely wrought much broader change \autocite{eli2025model}.} 
We conduct two exercises that reflect common empirical specifications. The first uses lags of the data to study the persistence of vehicle adoption across decades, while the second uses county fixed effects to test the relationship between population growth and vehicle adoption. The exercises are empirically taxing; dynamic panels and fixed effects can amplify the biasing effects of measurement error \autocite[e.g.,][]{griliches1986errors, wansbeek1992simple}. 
Across eight models (four models using distinct sets of the data for each exercise), hypothesis tests against a null of no effect using the LLM-based data yield identical conclusions to those derived from the manually validated gold standard data. Results reveal interesting patterns in vehicle adoption: strong persistence and a dynamic relationship with population growth over time. For six out of eight models, the key coefficients estimated from each data source are statistically indistinguishable. And when not indistinguishable, the results are informative for additional training that would improve model performance with a larger sample. 

This paper contributes to the evolving field of data digitization in five key ways. First, we design, implement, rigorously validate, and evaluate on a holdout sample a complete pipeline architecture leveraging multimodal LLMs specifically to create cohesive panel data from diverse historical table scans. This builds on prior work in both textual  extraction and table-specific extraction and extends it into the LLM era \autocite[e.g.,][]{shenLayoutParserUnifiedToolkit2021a, Dahl2023, Stelter2025}.\footnote{For a conceptually related application to textual patent data, see \textcite{griesshaber2025multimodal}.} 
Second, we demonstrate how this LLM-based approach makes large-scale table digitization both more cost efficient and significantly more accessible by allowing researchers to leverage domain expertise via prompts rather than technical skill. Third, our pipeline integrates alignment (harmonization), helpful for handling the heterogeneous table formats, differing geographic units, and disparate data categories common in historical tables.

Fourth, we adapt and formalize the use of gold-standard data to improve the performance of LLM-based digitization pipelines.  Unlike traditional pipelines in which observed errors are translated into opaque technical parameters of OCR and layout parsing systems, domain experts can identify LLM error patterns in natural language and directly adjust prompts accordingly. For example, observing empty cells being filled with hallucinated numbers led to adding the instruction ``[u]se \verb|""| for empty cells,'' while domain-specific knowledge that automobile counts are often labeled as ``passenger cars'', ``pleasure cars'', or ``owners'' led to adjusting the prompt to acknowledge that ``[column] matches are not always textually very similar.''
This domain-knowledge-driven refinement, conducted through natural language, makes pipeline development accessible to individual researchers rather than requiring specialized teams or acquisition of OCR and layout parsing expertise.

Fifth, while gold-standard evaluation is widely used in OCR and extraction studies \autocite[e.g.,][]{carlsonEfficientOCRBuilding2024, Gobel2013, Correia2023}, we extend this paradigm by performing a comprehensive \emph{end-to-end} evaluation proceeding from extraction to downstream econometric analysis, explicitly testing equivalence of economic inferences—vital for establishing trustworthiness of AI-processed data given potential error propagation concerns \citep{battaglia2024inferenceregressionvariablesgenerated}.

Reducing data digitization frictions is crucial because historical data are central to research, allowing scholars to trace the long-run evolution of social phenomena and understand their contemporary implications. From studying how the Dust Bowl shaped agricultural adaptation \citep{hornbeckEnduringImpactAmerican2012} to examining the lasting effects of social connectedness on crime \citep{stuartEffectSocialConnectedness2021}, or investigating how trade agreements influenced political realignment \citep{choiLocalEconomicPolitical2024}, historical data provide crucial insights into current economic and social conditions. Large-scale historical datasets, such as county-level vital statistics \citep{baileyUSCountyLevelNatality2016}, demographic records \citep{hainesHistoricalDemographicEconomic2005}, and environmental data \citep{gutmannGreatPlainsPopulation2005}, enable researchers to analyze long-term patterns and identify relationships that are difficult to establish using only contemporary data \citep{combes2022urban}. Indeed, reflecting sustained interest, grants related to historical tables from the NSF for Economics alone have totaled approximately \$46 million since 2000.\footnote{Based on NSF Award Search (SES Economics program, keywords "historical," "tables," 2000-present, accessed May 1, 2025).} Despite such investment, much crucial historical information remains locked in archival documents due to the extensive resources required for conversion into machine-readable formats suitable for analysis.

However, it is important to emphasize that these new tools are only valuable insofar as researchers can assess where these tools may introduce errors.  Such risks are not unique to LLMs: standard OCR, deep-learning layout models, and even manually entered data can all suffer from errors in table structure, numerical transcription, or harmonization. We therefore view creation of gold standard data and evaluation of the pipeline using such data as key ingredients for the success of our (or any) pipeline. 
We provide code, data, and practical guidance to help researchers adapt the workflow and expand the historical sources available for quantitative analysis \citep{abramitzky2025new}.\footnote{Our code and data are available at XXX (to be released after acceptance and prior to publication.}


\subsubsection*{Related Literature}

The research on automated extraction from tables consists of  three broadly related areas: pre-deep-learning pipelines, deep-learning pipelines, and prompt-based extraction with multimodal LLMs \citep{Singh2025, Fleischhacker2025}. 
Pre–deep-learning models are compared in \cite{Gobel2013}. Deep-learning OCR engines and layout models improve accuracy but typically require nontrivial parameter tuning and pipeline engineering.\footnote{One antecedent, \cite{Correia2023}, emphasizes that out-of-the-box deep learning OCR and layout parsing tools are insufficient and implements an iteratively improved pipeline using human-reviewed ground truth to tune the model parameters. In contrast, our approach replaces technical parameter tuning with prompt tuning guided by domain expertise and explicitly incorporates harmonization, yielding an \emph{end-to-end} pipeline that covers layout recognition, cell extraction, semantic normalization, and panel construction.} 
More recently, researchers have turned to multimodal LLMs to enable prompt-based extraction across domains, lowering technical barriers and often reducing cost \citep{Balsiger2024,Circi2024,Humphries2024,McLean2024,Stelter2025}.
Related studies applying deep learning or LLMs to historical documents \citep{Dahl2023,McLean2024,Stelter2025,Balsiger2024,Circi2024,Humphries2024} generally emphasize out-of-the-box extraction rather than a formal, domain-informed improvement loop that culminates in harmonized panels. This distinction is central for heterogeneous historical tables, where layout errors and harmonization choices can compound and ultimately affect downstream inference; our evaluation framework in Sections~\ref{sec:gs} and \ref{sec:auto} directly addresses this concern.

\section{Multimodal LLMs for Historical Table Digitization}

Multimodal LLMs differ significantly from traditional OCR methods, which typically handle layout and text extraction in separate, often brittle, stages. Multimodal LLMs integrate vision and language understanding, allowing for a more holistic interpretation of complex document structures, like historical tables with varied layouts and imperfections. This integration also permits guiding digitization using natural language prompts based on domain expertise, unlike the specialized coding needed for traditional tools.

Foundational technologies include the transformer architecture, adapted for both vision and language \citep{vaswani2023attentionneed, dosovitskiy2021imageworth16x16words, radford2021learningtransferablevisualmodels}. These models typically convert text and images into shared numerical representations (embeddings), enabling reasoning across modalities---connecting, for instance, a table cell's content with its header and visual position. While architectural specifics vary, their common strength is unified processing of text along with visual document features. See the discussion in Appendix~\ref{app:llm_technical} for more background and detail.

While general-purpose benchmarks for multimodal LLMs are evolving \autocite{liuOCRBenchHiddenMystery2024, fu2024ocrbenchv2improvedbenchmark}, there is an ongoing need for comprehensive benchmarks tailored to specific tasks like digitizing heterogeneous historical tables. Our paper addresses this need by evaluating LLMs on real-world historical economic tables, providing practical insights into their effectiveness at developing panel data for historical economic analysis.

\section{Data} \label{sec:data}

The primary dataset comprises publicly available scans of early 20th-century tables from within the U.S.\ recording annual, county-level vehicle registrations (including vehicle types like cars, trucks, etc.). This printed source material was produced by various state agencies (e.g., Departments of Transportation or Motor Vehicles), leading to substantial layout and content heterogeneity. Moreover, because some states have large numbers of counties and because many reports contain less relevant accounting data (e.g., fee revenue), many tables span multiple pages and contain extraneous and confounding information. This heterogeneity is relatively common with historical data and poses challenges for traditional OCR and layout parsing.

We require high-quality (``gold-standard'') data to first develop and validate the prompts we use to interact with the LLMs, and then separately to evaluate overall pipeline performance.\footnote{See Appendix \ref{sec:gs_evaluation} for additional details about creating and evaluating this dataset.} 
The gold standard data consist of 364 tables, which we then randomly split into separate subsets that we use for prompt development  and for overall evaluation (79 and 285 tables, respectively). While creating a gold standard dataset is labor intensive, it is \emph{necessary} for rigorous validation of a data extraction pipeline. Although we created the gold standard dataset upfront, in practice researchers can create this data incrementally during prompt development. Small batches of corrected LLM outputs can inform prompt refinements, creating a virtuous cycle that reduces subsequent manual effort.\footnote{Our methodology in creating the gold-standard data is not particularly novel; we do not emphasize this as a contribution. Rather, it is the availability of gold-standard data for prompt refinement and evaluation that is key.}

We apply the LLM-based pipeline to the same scanned images as in the gold standard data. This generates an ``LLM dataset,'' which we compare quantitatively to the gold standard dataset to evaluate digitization accuracy. Additionally, we employ the LLM dataset in our empirical analysis, verifying that conclusions drawn remain consistent with those derived using the gold standard data.\footnote{These tables are a subset of a larger effort to comprehensively catalog and digitize historical vehicle registration data. The gold standard data consist of most documents found as of the end of 2022, when processing began.}

\section{Historical Table Digitization Pipeline}

We create a multi-stage pipeline that converts scans of diverse historical tables into a cohesive panel dataset, using LLMs for table structure analysis, content extraction, and harmonization. We first describe the pipeline in Section~\ref{sec:pipeline_overview}. This pipeline can be adapted for use with other datasets. For source material similar to ours, a researcher can reuse our code and stage structure, reaching high performance through the iterative prompt development that we discuss in Section~\ref{sec:prompt_engineering}. For source materials that diverge, Section~\ref{sec:generality} discusses the design principles behind pipeline development, with our pipeline as an example. Though our main contribution is the pipeline itself, conveying these principles makes that contribution more transferable.

\subsection{Pipeline Overview}\label{sec:pipeline_overview}

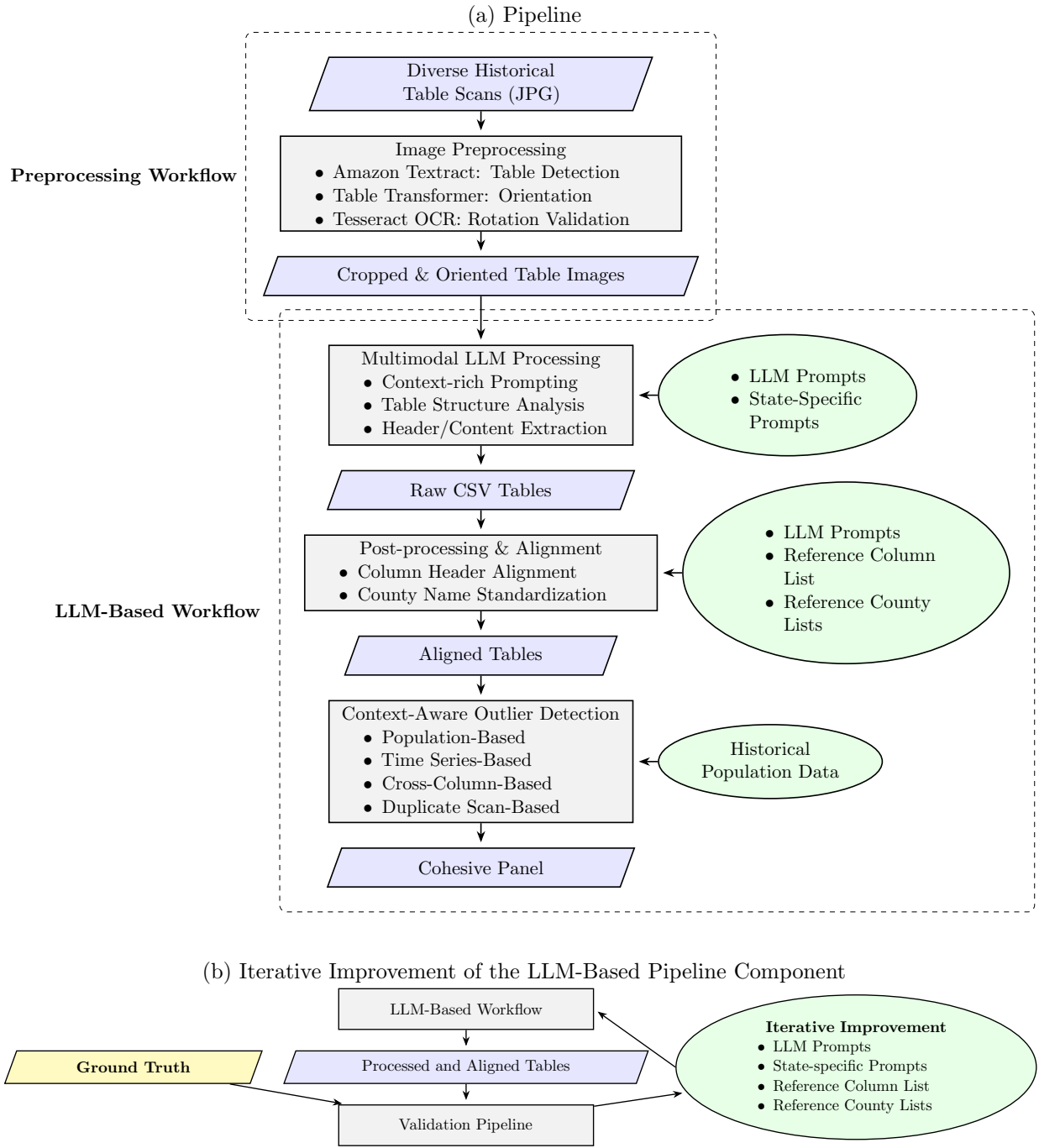
\begin{figure}[htbp]
    \centering
    
    \begin{subfigure}{\linewidth}
        \caption{Pipeline}
        \centering
        \resizebox{\linewidth}{!}{%
            \input{Figures/pipeline/pipeline}
        }

        \label{fig:pipeline}
    \end{subfigure}
    
    \vspace{0.5cm}

    \begin{subfigure}{\linewidth}
    \caption{Iterative Improvement of the LLM-Based Pipeline Component}
        \centering
        \resizebox{\linewidth}{!}{%
            \input{Figures/pipeline/loop}
        }
        
        \label{fig:iterative-loop}
    \end{subfigure}

    \caption{Historical Table Processing Pipeline}
    
\smallskip
\footnotesize \raggedright
    This figure illustrates the multi-stage historical table processing pipeline (a) and its iterative improvement cycle (b). In panel (a), the light blue parallelograms represent the data inputs and outputs at each stage, gray rectangles depict the processing steps, and green ellipses denote the \emph{user-provided inputs} that guide each step: the domain-specific knowledge we supply (LLM instructions: general, format-specific (county- vs.\ year-sorted), and state-specific; reference lists; and standardized categories) together with the tools the pipeline relies on (large language models and OCR / layout-detection models). Panel (b) shows how the LLM-based workflow is iteratively refined: its output is validated against ground-truth gold-standard data (yellow parallelogram), and the resulting diagnostics improve these inputs (green ellipses).
\end{figure}

\autoref{fig:pipeline} depicts the distinct stages with inputs and outputs at each step. The Image Preprocessing stage takes \emph{Source Table Images} as input and uses pretrained OCR and layout-detection models \citep{Smock_2022_CVPR, TessOverview} to detect each page's orientation and rotate it upright. These established tools require minimal development effort. This outputs \emph{Oriented Table Images}.

The pipeline then switches to an LLM-based workflow. The \emph{Oriented Table Images} enter the Multimodal LLM Processing stage, where domain knowledge-based prompts guide the LLM in analyzing table structure and extracting content. We develop prompts that first instruct the LLM to act as a careful researcher and avoid hallucinating numbers. Building on this, we add detailed instructions, for instance, to handle formatting conventions like commas, explicitly record empty cells, and recognize multi-row headers. Multi-page tables are submitted together as a single grouped set of pages so the model sees the full table at once.\footnote{We request JSON-formatted output adhering to a schema that we provide, and enforce this by selecting the structured output option in Gemini's API. Using an otherwise identical prompt, replacing the schema-enforced JSON request with a prompt-only request for CSV-formatted output in a 20\% subsample of our holdout data increases the total error rate from 7.7\% to 17.9\%, almost entirely by increasing the share of missing outputs from 6.9\% to 16.2\%.}

Beyond numerical transcription, one element of this stage is initial \emph{harmonization} of field and geography labels.  The model maps raw fields (usually column headers) to a predefined list of standardized categories (e.g., transforming ``Pleasure Cars'' or ``Passenger Cars'' to ``Automobiles''). While creating the standardized category list is specific to our domain, using the LLM for this mapping is generalizable and accessible. We augment this step with state-specific prompts that address unique reporting formats and geographic entities (which were often discovered during development). For example, we instruct the model that when Illinois data split Cook County into ``Chicago'' and the remainder, it should label these as distinct entities (``Chicago'' and ``Cook Excluding Chicago''). We detail and discuss the development of these prompts in Section \ref{sec:prompt_engineering}. Using structured outputs, this stage returns \emph{Structured Tables}.\footnote{After each extraction call, we check that the structured output has as many data rows as county labels. If not, we re-issue the identical call up to three more times and record LLM output as empty if these calls also fail.}

The Post-Processing \& Alignment stage refines the alignment  of the heterogeneously structured source tables into a homogeneous panel dataset---a common challenge for researchers. This step has two component tasks. The first structured prompt guides an LLM to standardize county names against reference lists, as well as flag non-county entities.  The second is further alignment of data category labels (column headers). Although column headers are already standardized during Multimodal LLM Processing, this task assesses differences in the standardized categories and gold standard categories. This is particularly helpful because determining whether category labels reference overlapping fields (e.g., types of vehicles) requires substantial judgment.\footnote{Because we are less certain of the validity of the human-assigned standardized field labels in the gold standard data, we evaluate these separately in Section \ref{sec:performance}.}
Although defining the reference category and geography lists is domain-specific, using an LLM for the mapping is generalizable and accessible. The output is \emph{Aligned Tables} with standardized fields and entities.

The \emph{Aligned Tables} compose an evaluation-ready, cohesive panel dataset. Domain-specific knowledge guides the pipeline. As shown in the green elements of \autoref{fig:pipeline}, we incorporate specialized expertise through carefully crafted LLM prompts and structured reference data (e.g., standardized column/vehicle-category list). These knowledge components emerge, in part, from the iterative refinement process: by comparing pipeline outputs against gold standard data, researchers identify key error patterns and articulate the domain knowledge needed to address them. This approach economically focuses human attention on the error patterns that matter most while effectively scaling domain expertise, as the developed knowledge components can then be applied to very large corpora of tables that would be infeasible to process manually.

\subsection{Iterative LLM Prompt Improvement} 
\label{sec:prompt_engineering}

The LLM-based pipeline uses iterative prompt improvement to achieve high accuracy  (as discussed above and illustrated in \autoref{fig:iterative-loop}). 
The cornerstone of the prompt refinement process is validation against the development subset of the gold standard data. 
Prompt refinement consists of two broad stages.

Initial prompts are exploratory with the target of developing a brief prompt that performs reasonably well across a broad range of settings.\footnote{This stage need not exist in every pipeline. Our use of the exploratory stage is akin to searching for a region of parameter space the broadly encompasses the optimum in maximum likelihood problems, whereas a more formal second stage corresponds to precise optimization.} 
As an example, we provide the initial prompt used at the beginning of our exploratory analysis below.
\begin{quote}
    {\bf Initial Exploratory Prompt}
    
    You are a researcher who carefully digitizes historical statistical tables. You look at scans from old books and put the tables in csv files. Output a csv table. Don't output any other text. The image to process is attached.
\end{quote}
Validation in the exploratory stage consists primarily of digitizing a table in the development data, then visually comparing the table with its gold standard analog. We then alter the prompt to address any observed issues, then re-digitize the table with the updated prompt. If the issue is addressed and no new problems are introduced, we move on to another table. If another problem is introduced, we again alter the prompt to explicitly address the problems. 

In our use case, this experimental stage led to the addition of several general purpose statements into the prompt. For example, we added the instruction  ``[u]se \verb|""| for empty cells''  to the LLM processing prompt after observing that empty cells sometimes contained extraneous symbols like dots. We also found that adding detailed structure on the relationship between the number of rows and columns helped reduce critical parsing errors (i.e., shifted rows and columns).

The second stage of prompt development is a structured workflow that prioritizes improving the prompt to fix observed errors while carefully validating performance so as not to reduce performance elsewhere.
To do so, we split the development data into 100 random subsamples and use it for development sequentially (i.e., starting with 10 subsamples, increasing to 20, and so on). At each stage, we analyze both aggregate and individual table errors. For aggregate performance, we generate summary statistics (such as those reported in Section \ref{sec:performance}) to track overall improvement. At the individual table level, we create difference tables that record discrepancies between the development data and LLM output. This helps separate critical parsing failures (e.g., shifted rows) from more isolated errors. We prioritize addressing errors with larger magnitude differences, as these have greater impact on downstream analysis.

This quantitative feedback informs prompt refinement. For example, as we experimented with longer and more detailed prompts that addressed formatting issues, we discovered that the LLM would often skip the second set of columns of tables that were wrapped within a single page. We therefore experimented with various iterations of the `wrapped-table rule' shown in the prompt below. After adding this instruction, we re-validated on the development set to confirm error reduction while ensuring that the new prompt did not inadvertently increase errors elsewhere. This approach capitalizes on researchers' understanding of their sources and context, allowing them to guide the digitization process without requiring specialized technical skills.

The final LLM-Based Workflow involves two calls to the LLM, each with its own goal and prompt, producing distinct intermediate outputs. This modular structure facilitates development and evaluation by allowing us to analyze and debug each component separately, and could be altered to address additional ingestion, transcription, or alignment issues. For instance, we can examine raw digitized tables before alignment, assess alignment quality independently, or insert additional processing steps where needed.

As an example, we provide the  final main prompt for the Multimodal LLM Processing stage of our workflow (elements in curly brackets are variables read into the prompt at execution). The full set of final prompts  and explanation of variables are listed in Appendix \ref{sec:app_prompts}.

\begin{quote}
    {\bf Main Prompt (final)} 
    
    You are extracting historical statistical tables from scanned images for academic research.

Your task is to find and extract ONLY tables that report county-level or city-level vehicle registration data from {state}.

Attached is \verb|{page_count}| from a historical document from \verb|{state}|.

Return JSON matching the provided schema exactly.

\verb|{format_instructions}|

Wrapped-table rule:
Some tables are wrapped to save space by repeating the same block of columns side-by-side, for example:

\verb%"County | Vehicles | County | Vehicles"%

If this occurs, unwrap the table:
\begin{itemize}[label=$-$]
    \item   treat each repeated block as the same schema
    \item append the rows from later blocks underneath the first block
    \item produce one long table with one set of \verb|column_names|
    \item do not duplicate headers in the output
\end{itemize}

If the same table is split across multiple pages, combine the continuation into the same table only when the table structure clearly matches.

Missing / unclear text:
\begin{itemize}[label=$-$]
    \item  It is very important to extract the exact text of the table. Pay careful attention to avoid mistakes.
    \item If a cell value is unreadable or genuinely blank, return \verb|""|.
    \item Do not guess. If a row label is unreadable, return the best transcription only if legible; otherwise return \verb|""|.
    \item Never infer numbers from surrounding totals.
\end{itemize}

Normalization rules:
\begin{itemize}[label=$-$]
    \item Remove commas from numbers
    \item  Remove dollar signs
    \item  Preserve decimals, minus signs, and parentheses if printed
    \item  Remove commas from county/city names and column names
    \item  Keep all values as strings
    \item  Use \verb|""| for empty cells
    \item  Do not use null, NA, or placeholders
\end{itemize}

Uniqueness rules:
\begin{itemize}[label=$-$]
    \item Column names must be unique within a table
\item Row names must be unique within a table
\item If printed labels repeat and clearly refer to distinct rows/columns, disambiguate minimally by appending a short printed qualifier rather than inventing content
\end{itemize}

\verb|{additional_instructions}|

Return only valid JSON matching the schema.

\end{quote}


\subsection{Pipeline Transferability and Design Principles}\label{sec:generality}

Although we expect this pipeline to transfer to a wide range of historical table collections similar to ours through the iterative prompt improvement described in Section~\ref{sec:prompt_engineering}, historical sources are heterogeneous and some will differ from ours in ways that updated prompts alone cannot bridge. A researcher may instead find their tables different enough to warrant rebuilding the pipeline by respecifying or expanding the sequence of stages in \autoref{fig:pipeline}. We describe how to proceed through a principled set of criteria that maximize the chance of producing trustworthy data.

We predicate pipeline design on three principles. The first is to describe the pipeline as a set of logically necessary stages and targets (what must happen to turn a scanned page into usable data) independently of how any stage is implemented, deferring the choice of implementation until later. The second is to use LLMs for extraction only within a \emph{low-autonomy} system. The third is to center both development and validation on comparison to a gold standard dataset. Together these constrain the set of admissible pipelines and give the researcher a framework for steering development.

The first principle separates \emph{what} the pipeline must do from \emph{how} each step is carried out. The logical components are the stages, each with its inputs, transformation, and outputs, together with the targets that follow from domain knowledge about what must happen to turn a scanned page into usable data. In our setting, the target is linked county-year-field information for each table, and the stages are detecting and isolating a table, extracting headers, rows, and cells, and aligning fields and entities; we exclude components that are irrelevant in our setting but may be central in others, such as handwriting extraction or language detection and translation.

The second principle is a broad implementation choice that governs how the LLM is used: it performs carefully scoped tasks \emph{within} fixed, predetermined steps rather than choosing its own course of action. This follows the now-standard distinction between \emph{workflows}, in which models and tools are orchestrated along predefined paths, and \emph{agents}, which choose their own sequence of actions.\footnote{For examples on best practices, see Anthropic's guidance on agent construction: Anthropic, ``Building Effective Agents,'' \url{https://www.anthropic.com/engineering/building-effective-agents}, last accessed June~9, 2026.} For our purposes, the deciding benefit of the workflow approach is more transparent verification. With fixed steps, each step can be checked in isolation against the gold standard. In contrast, an agentic model that directs its own process creates potentially as many behaviors to validate as there are paths it might take. Fixed steps also run faster, cost less, and give errors fewer opportunities to compound.

The third principle anchors pipeline development and evaluation on the gold standard. During development, it drives the direction of improvement. Every stage admits many possible implementations (e.g., rotation can be handled with classical tools or a model; header harmonization can be a dedicated call or, as in our pipeline, folded into extraction). As there is usually no \emph{ex ante} optimal choice, this principle dictates selecting empirically by enumerating candidates, forming hypotheses, and testing against the gold standard. The winning implementations are tuned to a project's own sources and model.\footnote{For prompts in particular, wording can be model-specific; this makes them the least transferable part of the pipeline and is why a new project should re-derive them against its own gold standard \citep{lu2022fantastically}.} 
Because the gold standard records column headers and individual cell values as separate quantities, it can be used to analyze each component separately. During evaluation, a holdout portion never used in development assesses the overall performance of the resulting pipeline. The gold standard is therefore not merely a measurement tool but an architecture decision that reflects the logical targets of the specification; we discuss this in Section~\ref{sec:gs}.

Looking forward, we expect a highly effective way to implement table digitization will be to provide our paper and code as context to an AI coding agent (e.g., Claude Code or Codex) and to develop the prompts or adjust the pipeline empirically on one's own document portfolio. Although the choice of design principles, especially the logical components and targets of extraction, is largely driven by domain expertise, AI agents can help the researcher enumerate the stages and ensure that the inputs and outputs of adjacent stages align. Once gold standard data exist, implementations for individual stages can be evaluated empirically. Within these constraints, the implementation details can be delegated to AI agents.

\subsection{Evaluating LLM Performance Using Gold Standard Data} \label{sec:gs} 

Rigorous evaluation of AI methods requires holdout data. While many researchers traditionally establish rigor by examining model internals, the opacity of many AI models, particularly closed-source LLMs, hinders such scrutiny. Evaluating outputs against holdout gold standard data that reflects the desired output allows for quantitative performance evaluation and aligns with broader scientific machine learning standards \citep{Kapoor2024}.

Gold standard data need not be costly. It can be created by manually correcting outputs from initial digitization attempts, using standard tools, custom software (as in our case), or even LLMs run with an initial exploratory prompt if every cell is carefully validated; Appendix \ref{sec:gs_evaluation} describes how we created ours. Reliable evaluation can often be achieved with moderately sized datasets, making this approach significantly more cost-effective than full-scale manual processing. There is no generally transferable answer to how large is large enough: a useful gold standard matches the distribution of the sources it will be used to evaluate and covers their major failure modes.\footnote{This mirrors broader best practice for evaluating machine learning systems, where evaluation data are judged by how well they represent the task at hand rather than by raw size. Our gold standard follows this principle, spanning the states, decades, and reporting formats in our sample.}
However, we train our pipeline on just 20\% of our data (79 tables), achieving strong  training sample performance (see Appendix Section \ref{sec:trainsample_perf}).  That we also achieve good out-of-sample performance suggests that gold standard data need not be very large in many applications.

Gold standard data may itself evolve during pipeline development. Like any transcription, it contains occasional errors, and iterative refinement often identifies them: when the pipeline and the gold standard disagree on a cell, we return to the original scan, and in some cases the scan shows that the recorded gold standard value, not the pipeline output, is wrong (for example, a mistyped digit or an inconsistently labeled split geography). In such cases, we then correct the gold standard to match the source document. Such corrections are small historical discoveries in their own right: the pipeline acts as a second, independent reading of the same page, and disagreement between the two readings localizes errors that neither would find alone.\footnote{A gold standard value is changed only when the original document shows it to be wrong. That is, the document (with human review) decides rather than the pipeline output, and the same rule applies whether or not the change favors the pipeline. To wit, prompts encode formatting rules and domain knowledge, never values from the gold standard, so at no point does the model observe the numbers it is evaluated against. Corrections therefore make the gold standard more faithful to the underlying records, sharpening evaluation.}

\section{Pipeline Performance}\label{sec:performance}

We evaluate the performance of the LLM-based digitization pipeline using the holdout (evaluation) subset of the gold standard data. This holdout subset is never used during prompt development.  We evaluate performance in three dimensions: whether a table is structurally usable, whether a cell can be matched to the corresponding validated cell, whether the matched numerical value is correct (or, in some cases, close to correct).
This approach allows some comparison with traditional layout parsing and OCR tools and lets us assess whether LLMs extract historical tabular data at a quality necessary for rigorous quantitative research.

For structural error analysis, we first introduce the concept of \emph{critical parsing errors} that capture the failure of the extraction process to produce a structurally sound and analytically useful table. This error occurs when an extracted table meets any of these conditions: (1) it contains no valid columns that can be used for numerical analysis, where a valid column is one where all cells contain only numeric values; (2) it has "extra cells," meaning some rows contain more content-bearing cells than there are columns defined in the header, indicating structural misalignment; or (3) the table is empty, containing zero rows of data.
Failing any condition marks the table extraction as invalid, since such structural issues render the data unavailable for subsequent analysis or require manual intervention to correct before the table could be meaningfully used.

Our LLM pipeline generates very few critical parsing errors. In the holdout sample, only one table (0.35\%) fails the critical parsing test. We compare the performance of this approach with  that of  Amazon Textract. We find substantial differences in critical parsing failure rates. As shown in \autoref{tab:critical_failures}, Textract fails on 61.4\% of holdout sample tables, rendering them structurally unusable without manual correction. 
This comparison captures a part of the digitization task that is distinct from pure numerical transcription and highlights the success of LLMs in capturing reasonable document structure.

The LLM-based pipeline also obtains high numerical accuracy. \autoref{tab:overall_metrics} reports the holdout-sample metrics for the pipeline run on Gemini 3.1 Pro.  
To assess numerical performance, we use Claude to align raw (unstandardized) field names, ensuring that pipeline-generated values can be compared with gold-standard values for the same table, county, year, and field.\footnote{We match on raw field names because we are less certain of the validity of the (human-assigned) standardized fields in the gold standard data. We later assess field alignment. Raw field matching performs very well: it is successful for all but one table, which then contributes to missing outputs in \autoref{tab:overall_metrics}.}
The total error rate across all extracted cells is 4.6\%. This is composed of two distinct failure types: missing outputs (1.4\%), where the pipeline produces no value for a cell the gold standard records, and incorrect outputs (3.3\%), where a value is produced but is numerically wrong. Manually examining the missing cells, we find that the large majority arise from rare, idiosyncratic table layouts that the non-LLM components of the pipeline are not built to handle; the rest are errors in matching raw column headers (about 102 cells, roughly 0.3\% of all extracted cells) or county names (2 cells) with the gold standard data. When a header or county cannot be confidently matched, the pipeline reports the cell output as missing. Such missing values are, in any case, straightforward to handle in practice---through imputation, alternative (perhaps manual) extraction, or further prompt improvement.

Among the 98.6\% of cells where the pipeline links a value, accuracy is high: 96.7\% match the gold standard exactly, the mean absolute percentage error is just 0.7\%, and 99.4\% of linked cells fall within 10\% of their validated value. Combined with the low rate of critical parsing errors, this places the extracted data at a quality level suitable for economic research, well beyond traditional layout-parsing and OCR-based alternatives. Among the 3.29\% of cells that are linked but incorrect (\autoref{tab:error_metrics}), the errors are mostly small---the median absolute percentage error is 5.5\%---though the distribution has a real tail (the 95th-percentile absolute percentage error is 25.4\%). Even on these incorrect cells the extracted values track the truth closely ($R^2$ of 99.43\%), indicating a modest number of errors rather than systematic bias. The data therefore support regression and aggregate analysis cleanly, while applications that turn on an exact value in every cell warrant further verification.

\begin{figure}
\centering
\begin{subfigure}{.48\linewidth}
\caption{Error Rates by Decade}
\includegraphics[width=\linewidth]{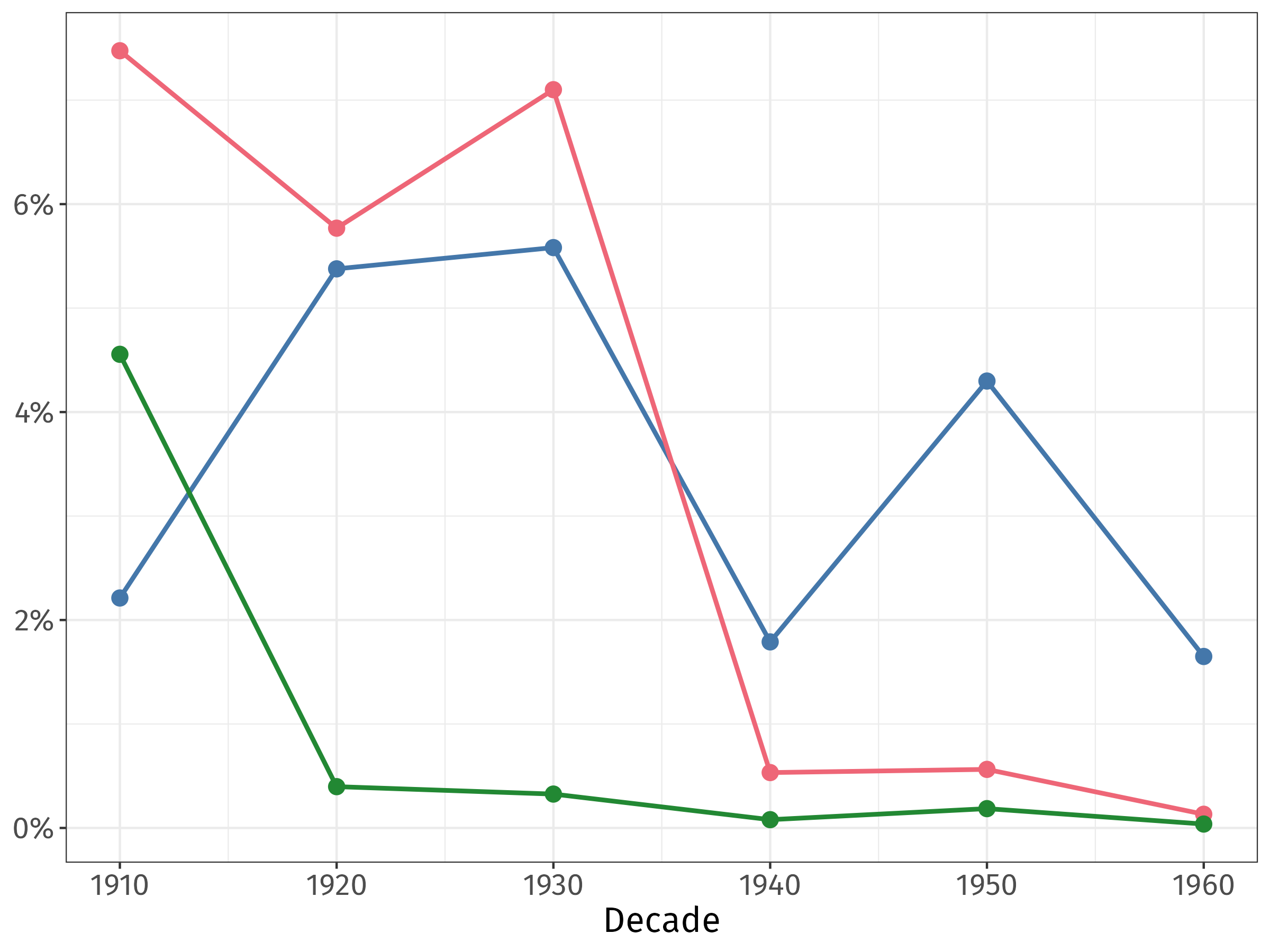}
\label{fig:error_decade}
\end{subfigure}
\hfill
\begin{subfigure}{.48\linewidth}
\caption{Error Rates by Geography}
\includegraphics[width=\linewidth]{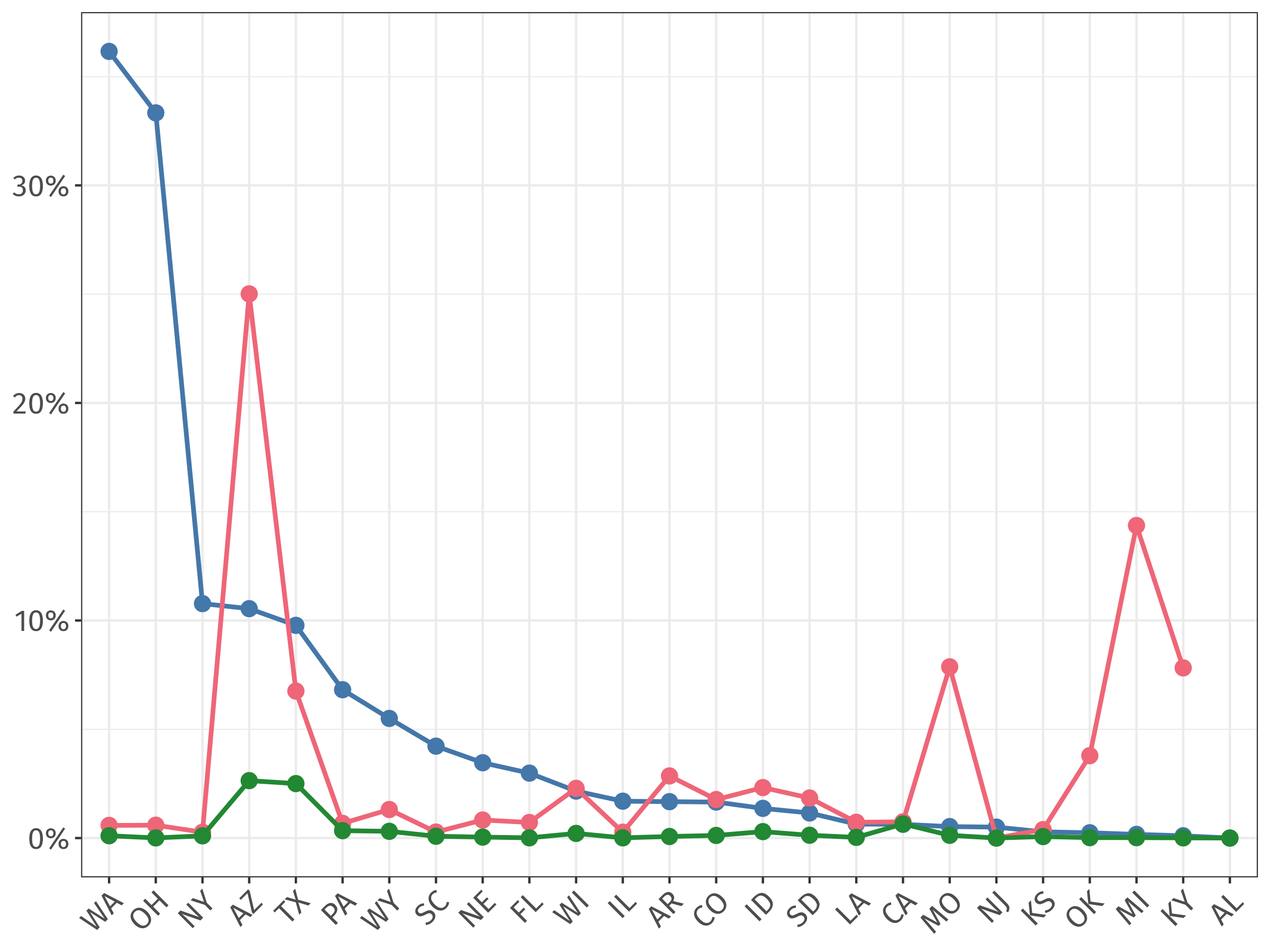}
\label{fig:error_state}
\end{subfigure}

\par\smallskip
\includegraphics[width=.8\linewidth]{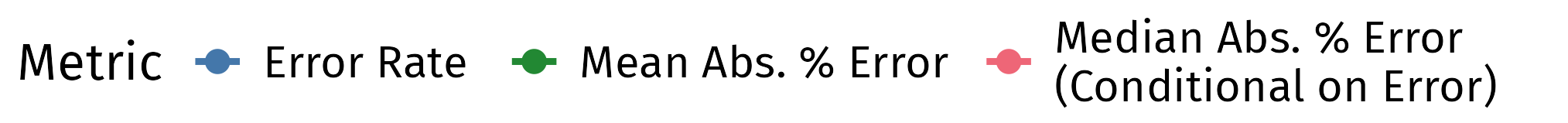}
\par\smallskip

\begin{subfigure}{.48\linewidth}
\caption{$R^2$ by Decade}
\includegraphics[width=\linewidth]{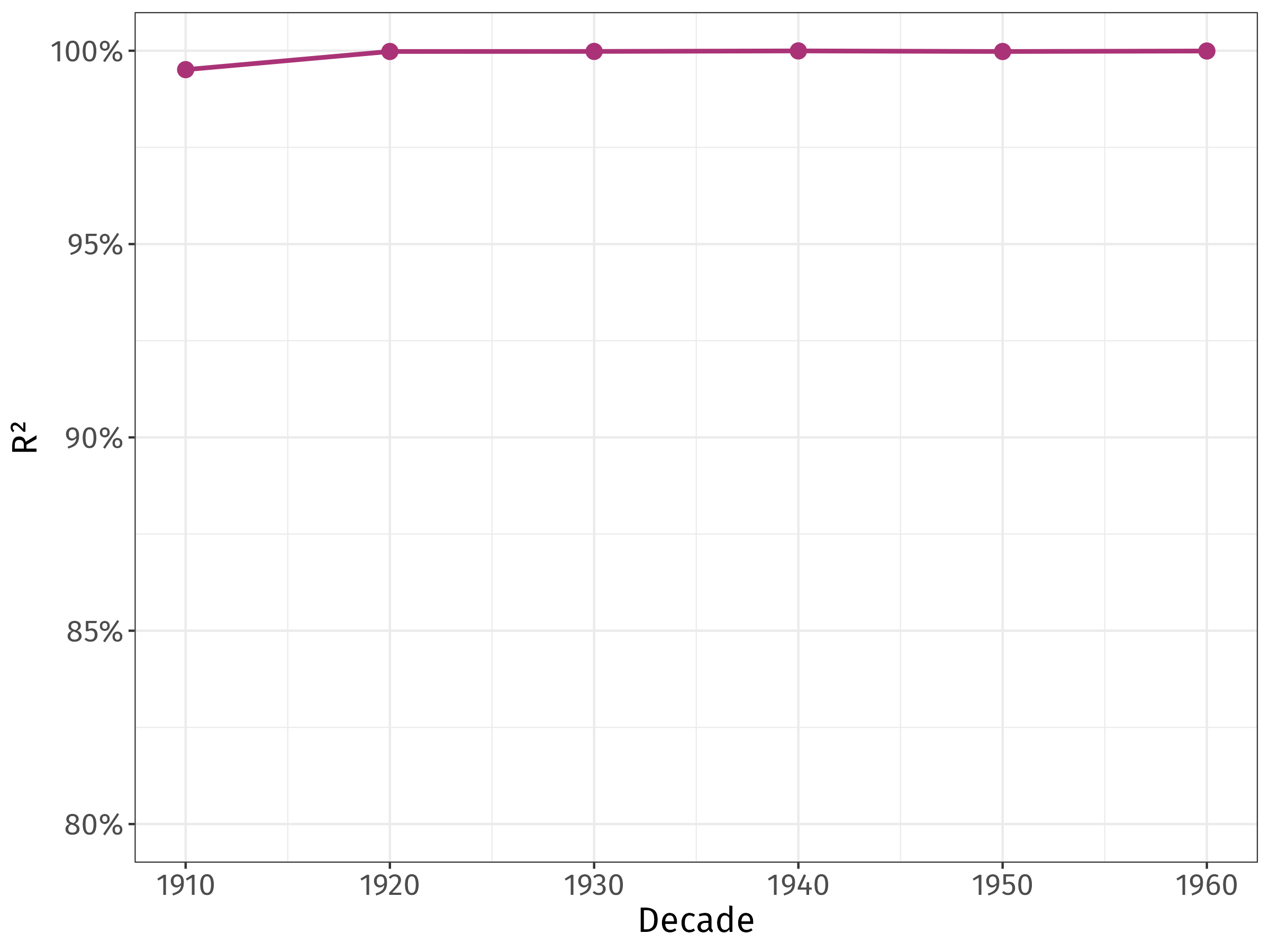}
\label{fig:r2_decade}
\end{subfigure}
\hfill
\begin{subfigure}{.48\linewidth}
\caption{$R^2$ by Geography}
\includegraphics[width=\linewidth]{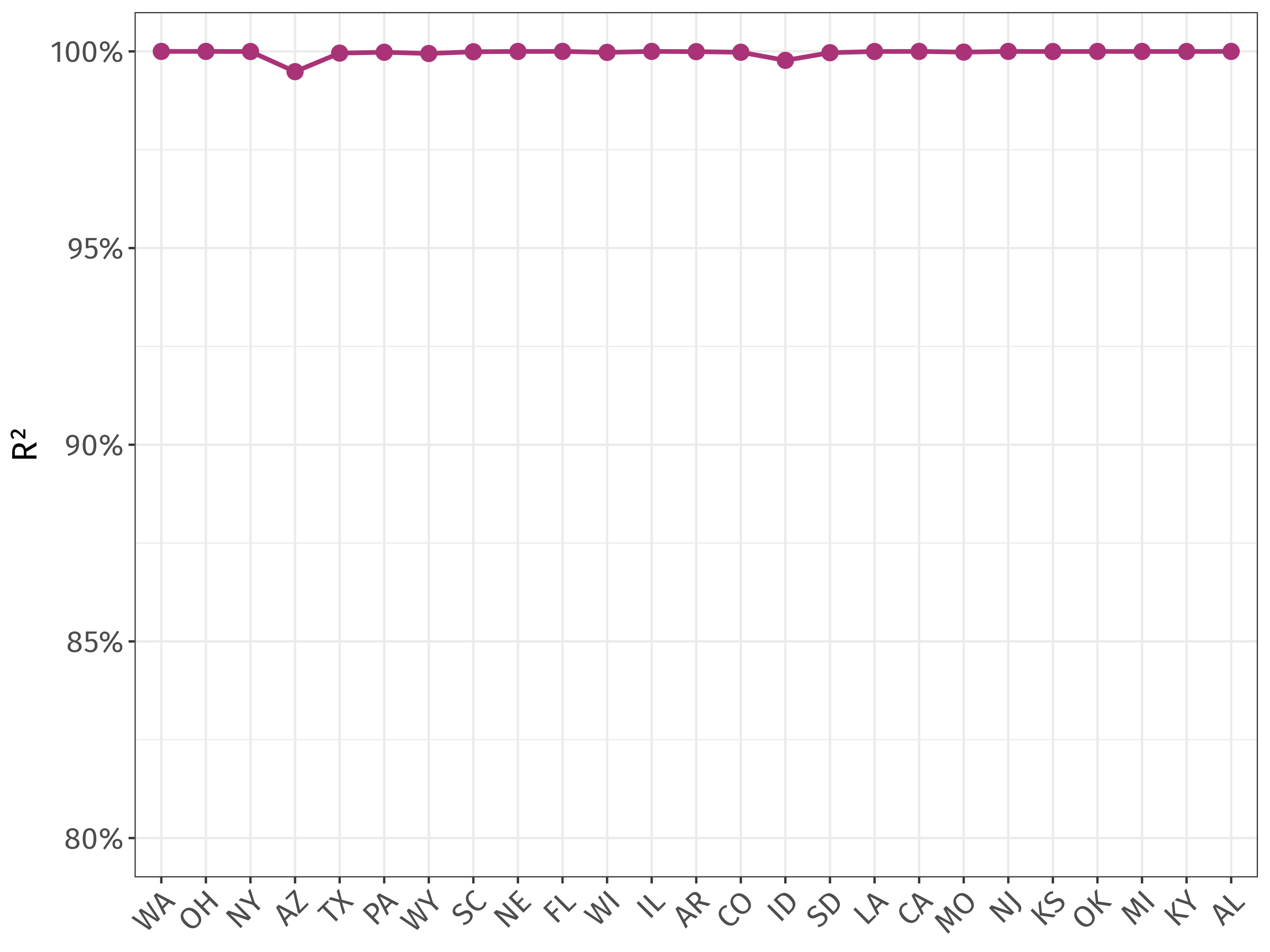}
\label{fig:r2_state}
\end{subfigure}
\caption{Performance Detail by Time and Geography}
\label{fig:performance_detail}
\smallskip
\footnotesize \raggedright
These plots examine the LLM pipeline's extraction quality consistency across heterogeneous historical tables from the holdout sample, assessing potential systematic performance variations linked to temporal (decades) and geographic (states) dimensions not evident from aggregate metrics. Error rate is total missing and incorrect outputs, mean absolute percentage error conditions on non-missing output, and median absolute percentage errors conditions on incorrect output (and is omitted if no cells have incorrect output).
\end{figure}

Since the dataset consists of heterogeneous tables that vary considerably across different states and decades, systematic errors could significantly affect subsequent analyses if performance degraded for particular subsets. \autoref{fig:performance_detail} examines extraction quality across time and geography to assess this concern. Results are reassuring: $R^2$ values remain consistently high across all decades (\autoref{fig:r2_decade}) and never fall below 99.5\%. Similarly, $R^2$ values remain above 99.4\% in every state (\autoref{fig:r2_state}). Error rates by decade (\autoref{fig:error_decade}) show the highest total error rate, approximately 9.5\% in the 1920s, while the median absolute percentage error for cells with errors remains well-bounded below 8\% across all decades. \autoref{fig:error_state} shows more volatility in error rates across states, but importantly, the highest state-level error rates are driven primarily by missing outputs concentrated in a small number of tables rather than by incorrect values. Even in states where error rates are higher, median absolute percentage error generally remains below 5\%. These patterns suggest that while performance does vary across contexts, the LLM approach maintains reasonable levels of accuracy across diverse tabular formats and historical periods.


\begin{table} 
\caption{Holdout Sample Performance (Textract and LLM Critical Parsing Failures)}
\label{tab:critical_failures}
\centering
\include{Tables/textract_comparison.tex}  
\smallskip
\footnotesize \raggedright
    This table compares LLM pipeline performance with Textract on the holdout sample, 285 tables not used in prompt development. Critical parsing failures are errors due to an extracted table (1) containing no valid columns, (2) having rows containing more content-bearing cells than there are columns (structural misalignment), or (3) having an empty table.
\end{table}

\begin{table} 
\caption{Holdout Sample Performance (Overall Performance Metrics)}
\label{tab:overall_metrics}
\centering
\include{Tables/overall_performance_test.tex}  
\smallskip
\footnotesize \raggedright
    This table quantifies LLM pipeline performance on the holdout sample (285 tables, 37,636 cells) not used in prompt development. Mean error rate, mean absolute error rate, mean percentage error, mean absolute percentage error, and $R^2$ are all conditional on a non-missing output cell match.
\end{table}

\begin{table} 
\caption{Holdout Sample Performance (Error Only Performance Metrics)}
\label{tab:error_metrics}
\centering
\include{Tables/error_only_performance_test.tex}  
\smallskip
\footnotesize \raggedright
    This table quantifies LLM pipeline performance on the holdout sample (285 tables, 37,636 cells) not used in prompt development. All statistics are  conditional on a non-missing but incorrect output cell match.
\end{table}


\paragraph{Header Alignment.} The pipeline returns 76 distinct raw field titles (headers), reflecting widely varying labels for different types of motor vehicle registrations and related concepts (e.g., registration fees). Performance statistics above reflect matching on these raw field names. However, the pipeline also proposes alignments of these raw field labels to a smaller set of categories (Header Standardization in Figure \ref{fig:pipeline}), which is useful to increase cohesion and facilitate downstream empirical analysis.\footnote{\label{fn:stdcats}The model is instructed to assign each vehicle column a standardized category: Automobiles, Trucks, Buses, Automobiles and Trucks, Automobiles and Buses, Trailers, Motorcycles, and Total Vehicles Registered, as well as a None type if not matched; the prompts listed in Appendix \label{sec:app_prompts} provide definitions for these categories.}
These standardized headers in the gold standard are based on subjective interpretation of field names in source tables. Because this categorization is partly subjective, we supply the standardized categories in the prompt but do not tune the prompt to reproduce the gold standard's assignments, instead treating the model's labels as an independent categorization. 
We then compare both the standardized headers in LLM pipeline data and in the gold standard data to an independent dataset of state totals, derived from Federal Highway Administration (Highway Statistics, table MV-201). This lets us assess how the LLM's category assignment compares with human-based category assignment, without conditioning on either being unambiguously correct.

Panel A of \autoref{tab:aligncoinc} reveals that coincidence with published state totals is broadly similar between LLM and human-based alignment. The human-based alignment in the gold standard data maps the 76 unstandardized headers into 8 categories, assigning a category to every cell. For the subset of data that can be aggregated into state totals, 88.3\% of state-field-year totals broadly coincide with those found in the published state totals.\footnote{This excludes tables that do not include all counties for a state. To facilitate comparison, we only compare Automobiles (with source fields `Automobiles' or `Automobiles and Buses') and `Totals' (with source fields `Totals' or `Automobiles and Trucks'. In practice, the number of buses is small compared to automobiles and trucks.}
The LLM pipeline maps the unstandardized headers into 13 categories, one of which is `none' and contains 1.2\% of cells.\footnote{Despite instructing the LLM to return one of the 8 categories listed in Footnote~\ref{fn:stdcats}, the LLM returned 12 distinct categories along with a `None' category.}
Coincidence with published state totals is only slightly lower than human-based assignment, at 85.8\%. 

Panel B directly compares the two datasets. It reveals 6 unique headers common across both datasets. Among all cells, 88.2\% of aligned categories coincide between the LLM and human generated datasets, where we again group conceptually similar categories. Examination of the cases where labels do not coincide often reveals ambiguous or conflicting source materials. For example, Trailers are sometimes a category in state tables and, when present, may be a constituent of a totals column. However, the published state totals in Federal Highway Administration (Highway Statistics, table MV-201) exclude trailers. Recognizing this, a researcher further developing the pipeline could train it to adjust totals in such cases to target coincidence with the published state totals.

\begin{table}[tp]
    \caption{Alignment Coincidence} 
    \label{tab:aligncoinc}
        \centering
\input{Tables/alignment}
\smallskip
\footnotesize \raggedright
    This table assesses the coincidence LLM pipeline header (field) alignment with human-provided labels. `Unst.' is unstandardized and `St.' is standardized. In Panel A, the denominator for column 3 is 50,653 (the total number of cells in the gold standard data). In Panel A, column 4 measures the number of state-by-field totals that coincide with the state totals published in Federal Highway Administration (Highway Statistics, table MV-201), conditional on the source material being amenable to calculating state totals (598 total state-field-year comparisons). In Panel B, header coincidence is a share of total cells (50,653).
\end{table}

\subsection{LLM-Based Pipeline Costs}

A significant advantage of the LLM-based digitization pipeline is the potential to substantially lower costs compared to traditional outsourcing solutions. This cost-effectiveness, combined with the simplicity of natural language prompting, positions LLM-based pipelines as a highly attractive alternative, democratizing high-quality digitization for a broad range of economic research applications.

Operational costs are primarily determined by the size of the text and image inputs processed by the LLMs and of the outputs they generate (in our case, Gemini 3.1 Pro for extraction and Claude Sonnet 4.6 for alignment). To establish a benchmark for comparison, we consider the costs of professional digitization services for similar historical tabular data. Normalizing LLM costs and digitization costs to our sample dataset reveals a striking average cost differential:
\begin{itemize}
    \item Small-scale outsourcing cost: \$8.24 per scanned page.
    \item Large-scale outsourcing cost: \$6.14 per scanned page.
    \item LLM-based pipeline cost: \$0.11 per scanned page (\$0.23 per table).\footnote{In our application, 4\% of LLM costs are associated with inputs (prompt text and scanned images) and 96\% with outputs, roughly 70\% of which reflects the model's reasoning tokens. Batch processing capabilities could further reduce costs.}
\end{itemize}
These figures indicate that the LLM-based approach is more than 50 times less expensive than outsourcing alternatives. This dramatic reduction makes large-scale digitization financially viable for many research teams.

While initial pipeline setup involves costs for iterative prompt refinement and gold standard creation, the latter can be made more efficient by using LLMs to produce initial gold standard drafts. This approach focuses the manual gold standard creation effort on correcting LLM outputs rather than generating data entirely from scratch, reducing labor costs. These overall setup costs are amortized over the project.\footnote{An exhaustive cost comparison involves many factors that are project (and institution) specific, such as whether gold standard data already exist, whether contracts or agreements with outsourcing providers are already in place, and labor costs. As an anecdote, the cost for a research assistant to digitize our data is roughly equal to that outsourcing. However, deploying an LLM-based pipeline would only require creating gold standard data for the portion of the overall source material used for training and validation (rather than the universe of data we use for evaluation).} 

\section{Empirical Assessment Examining Early Automobile Adoption}\label{sec:auto}

While section \ref{sec:performance} assesses whether the pipeline reads  historical tables accurately, high performance metrics do not necessarily ensure that analysis conducted on panel data compiled from the pipeline supports the same empirical conclusions as the gold standard data. We therefore conduct two exercises examining the persistence of automobile adoption between 1920 and 1960 and its relation to population growth in order to compare how the LLM data perform relative to the gold standard data. The first exercise regresses vehicle adoption on lagged vehicle adoption, while the second regresses vehicle adoption on population and includes county-level fixed effects. These two use cases can be unforgiving of measurement error: serial correlation amplifies bias in dynamic panels, and unit fixed effects amplify the attenuating effects of even classical measurement error \autocite[e.g.,][]{griliches1986errors, wansbeek1992simple}. Moreover, these exercises require reasonable field alignment. We do not make causal claims about the correlations we report, instead focusing on descriptive analysis and whether coefficients estimated with the LLM-processed data differ from those estimated with the gold standard data.

To proceed, we refine the \emph{Aligned Tables} into a format amenable to the econometric analysis in three steps. First, we only consider standardized fields that we can credibly align to match the concept of ``Total Vehicles'' (inclusive of trucks and buses).\footnote{We also manually address a few edge cases. In some years, California separates vehicles into fee exempt and fee paid categories, which we combine into total vehicles. We also manually assign ``Owner(s)'', ``Regular Number'', ``Total Cars'', ``Total Cars Including Exempts'', and ``Total Licenses Motor Vehicles" to ``Totals''.} 
These decisions reflect additional domain expertise we have built through working with these documents and the additional data increase sample sizes and the likelihood of detecting statistical differences between the data.
Second, because the data sometimes include multiple readings for each county-year-field cell, the data must be deduplicated and some geographies combined.\footnote{We avoid selecting on gold standard data to do so. When there are multiple candidate values, we average (ensemble) them. See Appendix~\ref{sec:duplicates} for full details.} 
Lastly, we conduct a minimal outlier detection procedure based on comparing the number of vehicles to county-level population; we exclude any counties where total vehicles exceed population by 50\%.

\begin{figure}[p!]
    \centering
        \includegraphics[width=0.95\linewidth]{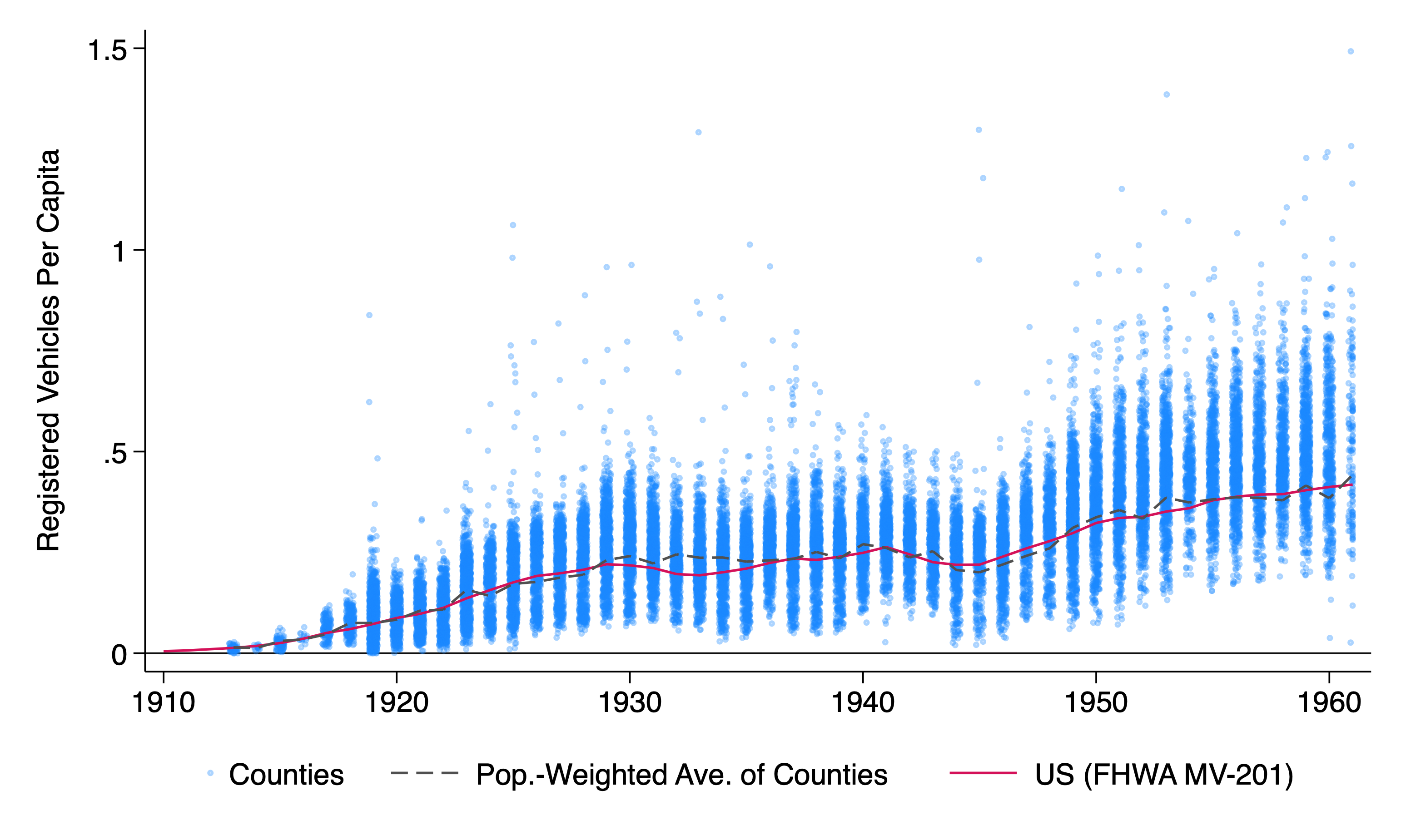}
            \caption{County-Level Vehicle Adoption Rates by Year, 1910--1961 (Cohesive Panel)} 
        \label{fig:counties}
 \fignote{This figure plots the distribution of data on vehicles per capita at the county level for each year in our sample. The dashed black line indicates the population-weighted average level of vehicles per capita in our dataset. The solid red line plots the reported national level of vehicles per capita from Federal Highway Administration data.}
\end{figure}

\autoref{fig:counties} plots the number of total vehicles per capita across counties and years, revealing substantial variation in the levels of vehicle adoption across counties. This figure also reports the population weighted mean of the LLM-based data and compares it to the mean vehicle adoption rate in the U.S. (from Federal Highway Administration, Highway Statistics, table MV-201). The mean LLM-based value closely tracks the national mean. This shows that the pipeline is reasonably accurate in aggregate and suggests that the observed sample of states is broadly representative of the nation. 

\subsection{Persistence in Vehicle Adoption} \label{sec:pers}

We first examine persistence in county-level rates over each decade from 1920 to 1960 by regressing the contemporaneous vehicle adoption rate on the vehicle adoption rate 10 years prior. We denote county-level log vehicle registrations per capita in county $c$ in state $s$ and year $t$ as $y_{cst}$ and estimate:
\begin{equation} \label{eq:pers}
y_{cst} = \rho y_{cs,t-10} + \delta_{st} + e_{cst},
\end{equation}
for $t \in \{1930, 1940, 1950, 1960\}$. To ensure this exercise has the power to statistically distinguish the data, we also include ten- or eleven-year differenced data within a one-year bandwidth of these years (so that, e.g., changes from 1921 to 1931 or from 1919 to 1930).\footnote{We only ever include one difference for each county for each $t$, prioritizing census years, then ten-year differences ending in 9's, then in 1's, then eleven-year differences.} 
Specifications also include state-by-year fixed effects, $\delta_{st}$, which play the dual role of isolating within state variation in adoption and controlling for measurement error that may be common to a source document or table or our processing thereof. We only include observations with both gold standard and LLM values.

In \autoref{eq:pers}, $\rho$ measures county-level persistence (serial correlation) in vehicle adoption. When $\rho$ is close to zero, there is little persistence in vehicle adoption between year $t-10$ and year $t$. When $\rho$ is close to one, persistence is very high across years. This parameter also measures  spatial convergence in adoption \autocite[e.g.,][]{barro1992convergence, mankiw1992contribution}.\footnote{The growth literature often expresses the correlation between contemporaneous and lagged values as a correlation between growth rates and initial values; in our notation, $y_{cst} = \rho y_{cs,t-10} \Leftrightarrow y_{cst} - y_{cs,t-10} = (\rho-1) y_{cs,t-10}$.}
For $\rho \in (0,1)$, growth is slower where initial adoption was higher, indicating convergence towards more similar adoption rates.  Adoption diverges if $\rho\geq 1$. Convergence is faster the closer $\rho$ is to 0. 
Examining $\rho$'s evolution reveals epochs of varying convergence or divergence.

We estimate \autoref{eq:pers} on two datasets: the LLM-based data and the gold standard data. 
The estimate that uses the processed LLM data we label as $\hat{\rho}^{\text{LLM}}$, whereas the estimate that uses the gold standard data we label $\hat{\rho}$. We stack these models to enable testing the null hypothesis of whether these coefficients are statistically distinguishable $H_0: \rho^{\text{LLM}} = \rho$. To account for within-county error correlation in both datasets, we cluster standard errors at the county level across datasets. 

Panel A of \autoref{tab:emp1} indicates substantial persistence in county-level vehicle adoption rates from 1920 to 1960. Estimates of $\rho$ vary from 0.26 to 0.79. Persistence was lowest earlier in the sample. Between 1920 and 1930, $\hat{\rho}^{\text{LLM}} = 0.26$ and $\hat{\rho} = 0.44$, indicating that increasing vehicles by 10\% more per capita in 1920 correlates with 3\%--4\% more vehicles in 1930, on average. 
Conversely, this means that convergence was greatest between 1920 and 1930; growth rates were somewhat slower in areas with greater early adoption. The 1920s were, in aggregate, an era of rapid adoption of vehicles \autocite{norton2011fighting}. The relatively low value of $\rho$ reflects broad-based increases in ownership that occurred in most US counties.

The following decades exhibited greater persistence in vehicle adoption. Estimates of $\rho$ after 1930 lie between 0.72 and 0.79, indicating that counties with 10\% higher adoption in one decade experienced 7\%--8\% greater adoption in the next decade. High persistence is perhaps not unsurprising  between 1930 and 1940, during which aggregate vehicle adoption rates flatlined (\autoref{fig:counties}).  \textcite{romer1990great} observes that the Great Crash reduced registrations, and while production fell sharply in 1930, reduced scrappage rates offset much of this decline \citep{chow1957demand}. Reduced investment and greater preservation of existing capital thus likely lay behind the higher local persistence and lower convergence across different counties seen in the 1930s.

The 1940s and 1950s also saw high local persistence in automobile adoption rates, with counties that had greater vehicle adoption maintaining relatively higher adoption levels than other counties, and vice versa. Historically, this period can be divided into two parts: While WWII reduced demand between 1941 and 1945 \autocite{flamm2006putting}, registrations rebounded post-war, growing rapidly through 1960. Unlike the more rapid convergence in vehicle adoption during the 1920s, aggregate growth in vehicle adoption of the 1940s and 1950s was substantially slower.

This persistence is long lived. Combining persistence coefficients across decades indicates that places with 10\% greater vehicle adoption in 1930 (1920) still experienced 4\% (1\%--2\%) greater adoption in 1960. As this time scale is much longer than the typical depreciation schedule of early automobiles, these results suggest that the initial factors that influenced early adoption have had long-lived influence \autocite{brooks2019vestiges}. The results also suggest that either people continually respond to those early differences \autocite[as in][]{severen2022formative}, and/or that factors complementary to vehicles adopted in some places are themselves quite persistent \autocite[e.g.,][]{bleakley2012portage, duranton2020economics}.

\begin{table}[tp]
    \centering
    \caption{Persistence and Correlates of Vehicle Adoption (LLM Data)} 
    \label{tab:emp1}
\resizebox{0.98\textwidth}{!}{
{
\input{Tables/newemp_1}

} }
\end{table}

For pipeline validation, we provide two points of comparison across the four pairs of models. First, all models reject the null hypothesis of $\rho=0$ and lead to the same economic conclusions: Automobile adoption was persistent between 1920 and 1960, and particularly so after 1930. 
Second, we test equivalence of $\hat{\rho}^{\text{LLM}}$ and $\hat{\rho}$. 
For three out of four of these tests, $\hat{\rho}^{\text{LLM}}$ is statistically indistinguishable from $\hat{\rho}$, reflecting the typical high accuracy of the LLM data documented in Section \ref{sec:performance}. For the 1920 to 1930 comparison, however, we reject $\hat{\rho}^{\text{LLM}} = \hat{\rho}$ at conventional levels. We do not interpret this as evidence to discard the LLM data.\footnote{Were this exercise solely focused on minimizing bias (rather than pipeline evaluation), additional techniques, such as design-based supervised learning \autocite[which uses gold standard data to improve precision as in][]{egami2023using} or prediction-powered inference \autocite{angelopoulos2023prediction}, would likely prove useful.}
In fact, examining the data to find the cause of this statistical difference may provide useful information for further pipeline improvement.\footnote{In this case, statistical inequivalence comes primarily from one source. The first is that the source table for Texas counties in 1919 features one page with a warped scan, and the LLM assigns numerical values to counties incorrectly. Interestingly, a prior version of the pipeline using earlier LLM models correctly assigned these values. This suggests even though newer models typically improve average performance, this does not imply that they will not  introduce new failure modes.}

County-level estimates of the dynamics of vehicle adoption have not been broadly estimable prior to this due to a lack of local panel data.\footnote{One exception is \textcite{meir1981innovation}, who studies vehicle adoption in Ohio counties in the 1930s.}
\textcite{eli2022transportation} estimate state-level persistence in vehicle registrations from 1919--1929, finding 10 fewer vehicles per 10,000 people in 1919 is correlated with 1.8 percentage point (pp) faster growth in vehicle adoption rates over the next decade. Although they use a different model than \autoref{eq:pers}, adapting their model to our data, we estimate that 10 fewer vehicles per 10,000 people in 1920 is correlated with 0.4--0.5 percentage point faster growth in vehicle adoption rates. 
Comparing estimates reveals that local convergence is roughly four times slower than state-level convergence. This suggests that the variation in adoption across counties within the same state is much greater than the variation in adoption across states. This underscores the value of our digitization pipeline in developing richer data to broaden economic conclusions.

\subsection{Population Growth and Vehicle Adoption}

The next empirical exercise examines the correlation between population growth and vehicle adoption and uses panel-unit (county) fixed effects. This type of specification is common in applied research, and so provides another test of pipeline performance. As before, we do not focus on establishing a causal relationship. Rather, our intent is to test whether economically interesting patterns are equivalent in both the LLM data and our gold standard data.

Specifically, we regress annual log vehicle registration rate per capita ($y_{cst}$) on log county population:
\begin{equation} \label{eq:pop}
    y_{cst} = \beta \ln(\text{pop}_{cst}) + \alpha_c + \delta_{st} + e_{cst}.
\end{equation}
This specification includes county fixed effects ($\alpha_c$) to ensure that $\beta$ reflects changes in population and vehicle adoption and to limit confounding factors that are time-invariant at the county level, like location-specific factors of production (e.g., ports). We estimate this model in ten-year increments to increase the importance of the county-level fixed effects (as they are frequently pivotal in applications); the power of unit fixed effects to control for time invariant factors decreases as panel length increases \autocite{millimet2023fixed}. As before, we include state-year fixed effects and only include observations for which we observe both gold standard and LLM values.

We limit the sample of counties to those with a population of less than 50,000 in the initial year of each decade in order to isolate the comparison between places that experience rapid growth with places that do not.\footnote{Note that annual population is linearly interpolated between decades. The same interpolation is used for both calculating per capita vehicle registrations and the independent variable.}
We estimate the model in \autoref{eq:pop} on both datasets, restricting the sample to include only observations present in both (as in Section \ref{sec:pers}). Estimates that use the processed data are labeled $\hat{\beta}^{\text{LLM}}$, whereas those from the gold standard data are denoted $\hat{\beta}$. We again stack models when testing the null hypothesis of $H_0: \beta^{\text{LLM}} = \beta$, clustering standard errors at the county level across years and models.

The parameter $\beta$ reflects the (correlational) elasticity of per capita vehicle adoption with population growth (i.e., a 1\% population change correlates with a $\beta$\% change in per capita vehicle adoption). Values of $\beta$ close to zero indicate that vehicles are being adopted as fast as the population is growing. For $\beta>0$, vehicle adoption is greater than population growth, whereas people adopt vehicles at a slower rate than population growth for $\beta<0$. The anticipated sign and magnitude of this correlation depends on how transportation technologies and patterns of land use coevolve in the face of population growth \autocite{cervero1997travel, bento2005effects, ewing2010travel}. If growth is accompanied by investment in non-automobile infrastructure and integrated land use patterns, automobile use may fall. However, if growth is served by roads and land use is segregated, private vehicles may become more necessary. And population growth may itself signify increased economic opportunity and thus income growth, which could increase vehicle adoption \autocite{dargay1999income}.

Panel B of \autoref{tab:emp1} shows intriguing variation in the dynamics of the relationship between population growth and vehicle adoption. The coefficient for the 1920--1930 window is positive, indicating that vehicle registration growth exceeded population growth in the 1920s---the counties with the greatest population growth were also those with the largest increase in automobiles per capita. This suggests that in the 1920s, urbanization and changes in transportation technology are likely intertwined. The coefficient becomes positive but is insignificant in the 1930s, indicating that population growth and vehicle adoption are in sync. Thus, during the Great Depression, urbanization was not systematically associated with changes in transportation technology.  This changes again after 1940. Between 1940 and 1950, a 10\% increase in population is accompanied by a statistically significant 3\% decline in vehicle adoption per capita. In the following decade, the elasticity is smaller in magnitude but remains significant, indicating a 10\% increase in population occurs with a 2\% decline in vehicle adoption. 

Because \autoref{eq:pop} includes county fixed effects and county land area is fixed, these results can be interpreted as the elasticity between population density and vehicle adoption. \textcite{duranton2018urban} provide careful cross-sectional analysis of the relationship between density and vehicle miles traveled in modern times. They find an elasticity of about -0.07. Our estimates are somewhat different in nature, reflecting historical periods and exploiting changes in population over time within county. Nonetheless, we find estimates both smaller and greater in magnitude than those in \textcite{duranton2018urban}, reflecting different periods of growth and urbanization.

Panel B also provides the second validation exercise using the LLM data. As before, all four models come to the same economic conclusion regarding the null hypothesis of $\beta=0$ and lead to the same qualitative conclusions. And in three of four of these tests, $\hat{\beta}^{\text{LLM}}$ is statistically indistinguishable from $\hat{\beta}$. However, for the 1940--1950 window, we reject $\hat{\beta}^{\text{LLM}} = \hat{\beta}$ at the 10\% level. Despite this, the numerical difference in the estimated elasticities is small; $\hat{\beta}^{\text{LLM}}$ is less than 3\% smaller in magnitude than $\hat{\beta}$.

\section{Conclusion}

For printed historical tables like the vehicle-registration records studied here, multimodal LLMs prove an effective and relatively inexpensive tool to translate heterogeneous scans into panel data. The LLM-based pipeline architecture we develop and evaluate drastically reduces critical parsing errors compared to standard tools. This approach allows researchers to leverage domain expertise via natural language, lowering technical barriers. The resulting county-level vehicle adoption dataset reveals granular dynamics obscured by state-level data, demonstrating the approach's potential. By dramatically reducing data acquisition costs, such LLM-based methods can fundamentally shift the optimization calculus for data-hungry researchers, enabling research driven more by potential insight than by data accessibility constraints.

As these tools lower the cost of digitizing diverse historical sources, rigorous validation and evaluation become paramount to ensure data reliability. We advocate for accompanying AI-generated datasets with transparent evaluation, using methods like the gold standard approach, to maintain the integrity of research findings. The broader lesson is not that LLM-generated data be accepted without scrutiny, but that using LLMs to generate historical data can and should utilize researchers' extensive domain expertise, validation against gold-standard data, and (when possible) downstream empirical validation to ensure reliable data.


\newpage
\singlespacing
\printbibliography

\clearpage 
\processdelayedfloats
\clearpage


\makeatletter
\efloat@restorefloats
\makeatother

\newpage
\appendix

\pagenumbering{gobble}
\pagenumbering{arabic}
\renewcommand*{\thepage}{A-\arabic{page}}

\setcounter{table}{0}
\renewcommand{\thetable}{A\arabic{table}}
\setcounter{figure}{0}
\renewcommand{\thefigure}{A\arabic{figure}}
\setcounter{equation}{0}
\renewcommand{\theequation}{A\arabic{equation}}

\section{Supplemental Appendix}

\subsection{Multimodal LLM Technical Background}
\label{app:llm_technical}

The multimodal LLMs, such as Anthropic's Claude 3.5 Sonnet and Google's Gemini 1.5 Pro used in our pipeline, build upon several key technological components. Text processing relies on transformer-based language models \citep{vaswani2023attentionneed}. Input text is converted into numerical tokens via a tokenizer, and these tokens are then mapped to dense vector embeddings that capture semantic meaning.

Visual information is processed by specialized vision models, often adapted from architectures like Vision Transformer (ViT) \citep{dosovitskiy2021imageworth16x16words}. These models transform pixel data into numerical representations that encode features ranging from broad structure (like table layouts) to fine details (like cell boundaries and typography). Some models use hybrid encoders processing images at multiple resolutions to capture both aspects effectively.

A critical step is aligning the representations from the text and image modalities. As an example, consider the DeepSeek-VL model \citep{lu2024deepseekvlrealworldvisionlanguageunderstanding}. While not used in our final pipeline, its architecture illustrates common principles. Text is processed by a language model backbone that converts words into numerical tokens via a tokenizer, then transforms these tokens into embeddings---dense vectors capturing semantic meaning where similar words cluster together. Visual inputs are handled by a hybrid encoder system that processes images at different resolutions, transforming pixel data into numerical representations that preserve both broad content (table structure) and fine details (cell boundaries, typography). Text and image representations are then aligned through a two-layer adaptor consisting of multilayer perceptrons---essentially a set of regression-like functions where each input element influences every output element through learned weights with non-linear transformations. This adaptor maps the different visual features into the same dimensional space as the language embeddings, creating a unified numerical framework where both text and image information become compatible vector representations. This mathematical alignment enables the model to effectively process tables by preserving the critical relationships across modalities---connecting the meaning of content with its position in the table grid.

More generally, alignment is often achieved using adapter layers or cross-attention mechanisms. These components project visual features into the same high-dimensional vector space as the text embeddings, creating a unified framework where the model can jointly reason about visual layout and textual content, preserving crucial relationships like the association between a number and its corresponding row and column headers in a table.

Architectural approaches to integrating modalities vary. \cite{wadekar2024evolutionmultimodalmodelarchitectures} categorize them into types such as deep fusion (modifying internal LLM layers for cross-modal attention) and early fusion (connecting separate text/vision encoders at the input stage, sometimes after tokenizing images). The specific models used in this paper fall broadly under these integration strategies, enabling the end-to-end processing described in Section~4. For deeper technical surveys and architectural details, see \cite{wadekar2024evolutionmultimodalmodelarchitectures} and the documentation for the specific models employed.

\subsection{Textract Limitations and LLM Advantages for Historical Table Processing}\label{sec:textract}

This appendix provides a concrete example of the key Amazon Textract Layout parsing errors that contribute to its high rates of critical failures discussed in Section \ref{sec:performance}. Using the 1923 Michigan vehicle registration data shown in \autoref{tab:mi_comparison}, we demonstrate how structural misinterpretations in OCR processing render extracted data unusable for downstream economic analysis. We then contrast these results with the significantly more accurate output generated by Large Language Models (LLMs) when processing identical source material.

Examining the Textract-extracted data in \autoref{tab:mi_textract_example}, we observe two key layout parsing failures relative to the input table scan in \autoref{tab:mi_original}. 
First, Textract incorrectly disjoints rows that should be unified. This is evident in the first data row, where Alcona County's data is split across multiple rows, with the county name separated from its corresponding values. The passenger car count for Alcona (733) appears in a row without a county identifier, while subsequent data shifts position relative to their proper county associations.
Second, Textract incorrectly joins values within cells that should be separate. For instance, in the Allegan row, the commercial vehicles count appears as "900 234" instead of the correct value of 909. Similar merging errors occur in the Alpena row, where "2671 1575" appears instead of distinct values in separate cells.

These two error types propagate to create additional structural problems: rows containing numeric data without county identifiers and counties without complete data values. The error in the first row (Alcona) is particularly destructive as it cascades throughout the entire table, causing misalignment between counties and their data points. This single parsing failure shifts all subsequent data values relative to their county identifiers, effectively rendering the entire dataset unusable for economic analysis without substantial manual intervention.

The LLM-structured output in \autoref{tab:mi_llm_example} demonstrates consistent accuracy in preserving the original table structure. The LLM correctly associates each county with its corresponding data across all columns, maintains proper row boundaries, and accurately distinguishes between adjacent numeric values. This structural coherence avoids the cascading errors observed in the OCR output and produces data suitable for immediate economic analysis.

LLMs are not without transcription errors, however. For example, for Baraga County, the LLM incorrectly reads the Trailers value as 2 when it is actually 0---a value correctly captured (as blank) by Textract. Nevertheless, these isolated numerical errors, which are limited in frequency (as we document in Section~\ref{sec:performance}), are dwarfed by the widespread layout parsing failures associated with Textract that render entire datasets unusable.

While layout parsing and OCR systems like Textract could potentially be optimized to reduce the errors, such optimization requires substantial technical expertise in document processing that is often orthogonal to researchers' core competencies. Each historical table format would likely require different OCR configurations, creating additional workflow complexity when processing diverse archival sources (the core objective of this paper).

In contrast, LLMs demonstrate robust table parsing capabilities with minimal specialized configuration. The high-quality results are achieved largely "out of the box," with additional refinements implemented through prompting techniques that leverage subject matter expertise rather than technical programming knowledge. When data inconsistencies do arise, researchers can address them using their domain knowledge of historical data patterns and structures, working within their established methodological frameworks rather than acquiring tangential technical skills.

\begin{table}
\caption{Michigan Vehicle Registration Data Comparison} 
\label{tab:mi_comparison}
\begin{subtable}[t]{0.48\textwidth}
\caption{Original Michigan 1923 Vehicle Data}
\label{tab:mi_original}
\centering
\includegraphics[width=\textwidth]{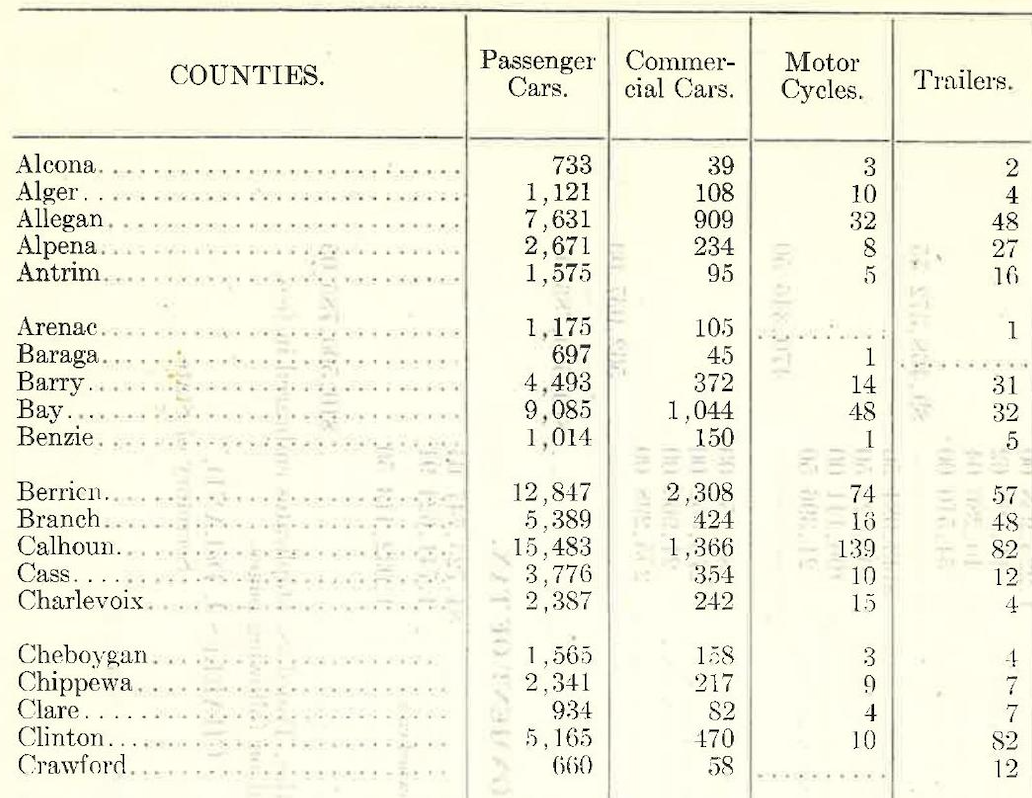}
\end{subtable}%
\hspace{0.01\textwidth}
\begin{subtable}{0.48\textwidth}
\caption{Textract-Extracted Vehicle Data}
\label{tab:mi_textract_example}
\centering
\resizebox{\textwidth}{!}{%
\input{Figures/comparison_with_textract/example_textract.tex}%
}
\end{subtable}%
\vspace{0.5cm}
\begin{subtable}[t]{0.48\textwidth}
\caption{LLM-Structured Vehicle Data}
\label{tab:mi_llm_example}
\centering
\resizebox{\textwidth}{!}{%
\input{Figures/comparison_with_textract/example_llm.tex}%
}
\end{subtable} 

\smallskip
\footnotesize \raggedright
This table visually contrasts the outputs from different digitization methods for the 1923 Michigan vehicle data. It presents the original document scan (a), the structurally flawed extraction by Amazon Textract (b) exhibiting issues like incorrect row splitting and merged cell values, and the corresponding structurally accurate extraction by the LLM (c), which, while not entirely free of numerical transcription errors (e.g., Baraga Trailers) preserves overall table integrity.
\end{table}

\subsection{Creation and  Evaluation of the Gold Standard Data}\label{sec:gs_evaluation}

We create the gold standard data using a combination of existing OCR tools (including Textract), custom software, and manual validation. The custom software juxtaposes scanned images with similarly formatted tables and enables rapid manual correction. Multiple passes through this manual digitization step increase accuracy, which we confirm through additional manual evaluation (see below). Gold standard data creation is labor intensive, but we have high confidence in the quality of this dataset.


To provide independent evaluation of the quality of the gold standard data, we randomly selected 100 data points from the gold standard data (10 documents with 10 data points per document) and asked a research assistant to record the values in these cells. Importantly, the research assistant did not observe cell values in the gold standard data. We then compare the cell values input by the RA with those in the gold standard data. 

The values precisely matched in 98 out of 100 cells. In the two cells that did not match, it was the gold standard data that were correct; the research assistant had incorrectly entered data from an adjacent cell. Thus, careful ex post evaluation reveals that the gold standard data is 100\% (in this small random sample). Moreover, this exercise highlights that manual data entry itself is, as always, subject to human fallibility.

\subsection{Training Sample Performance}\label{sec:trainsample_perf}

\autoref{tab:trainsample_perf} shows the same performance metrics presented in \autoref{tab:overall_metrics} but on the training sample. Within-sample performance is strong using Gemini 3.1 Pro, achieving a match rate of 97.5\% and, conditional on match, a median absolute percentage error of only 0.1\%. This in part led us to use only 20\% of the sample for training.

\begin{table}[htbp]
\caption{Training Sample Performance (Overall Performance Metrics)}
\label{tab:trainsample_perf}
\centering
\include{Tables/llm_metrics_fold1-20} 
\smallskip
\footnotesize \raggedright
    This table quantifies LLM pipeline performance on the training sample (79 tables, 12,700 cells) used for prompt development. Mean error rate, mean absolute error rate, mean percentage error, mean absolute percentage error, and $R^2$ are all conditional on a non-missing output cell match.
\end{table}

\subsection{Removing Duplicates}\label{sec:duplicates}

For a substantial share of county-by-year-by-field cells, the source material contains more than one reading for the same county-by-year concept. This occurs primarily for two reasons. First, sequential state reports sometimes reprint prior-year registration data, so the same county-year cell appears in more than one document. Second, historical reports do not always use stable vehicle-category labels: a raw heading such as ``owner,'' ``regular number,'' ``total cars,'' or a fee-paid/fee-exempt distinction may need to be translated into the categories used in the empirical panel. We must choose among these duplicates (or combine them) to proceed with the empirical analysis in Section \ref{sec:auto}.

The deduplication procedure separates duplicate raw readings from vehicle-category harmonization. We first identify records with the same county, year, and raw field label. Within these groups, we retain records that contain both the manually validated value and the LLM-extracted value when such paired records are available, remove exact repeated rows, and collapse (average) repeated readings that share the same raw field label and the same harmonized field assignments. This step treats multiple appearances of the same source cell as repeated measurements of the same observation rather than as independent county-year observations.

We then resolve duplicates that remain after field harmonization. When repeated records agree on both the manually validated field assignment and the LLM field assignment, we collapse them to a single value. A small number of state-specific cases require explicit treatment because the historical reports split or label categories in ways that do not correspond one-to-one to the final labels. Two primary cases are early-1950s California records with fee-paid and fee-exempt components (these are summed where the source structure implies a total) and Missouri records from the 1920s and 1930s with multiple readings under the same validated field (these are averaged and treated as totals).

Finally, we map the remaining labels into the common labels used in the empirical exercises: Automobiles, Automobiles and Buses, Automobiles and Trucks, and Total Vehicles Registered. Some raw labels are substantively totals even though they do not use the word ``total''; for example, labels such as ``owners,'' ``regular number,'' ``total cars,'' and ``total licenses motor vehicles'' are mapped to total vehicles.\footnote{In earlier documents, ``total cars'' sometimes refers to all types of vehicles, not just passenger vehicles.} 
We also make a small set of state-year corrections where the historical reporting convention is known to differ from the generic rule. The output preserves separate variables for the manually validated value and the LLM-extracted value, allowing the downstream analysis to compare the revised pipeline panel with the gold-standard panel on the same county-year-category structure.

This de-duplication procedure does not choose among conflicting readings by selecting the value closest to the eventual regression result, the state total, or a county's per-capita vehicle rate.

\subsubsection{Combining Geographies}

The empirical exercise combines the following geographies for consistency and comparability over time, as source documents often but inconsistently separate out large cities from their encompassing county. Outside of this list, geographic entities are nominal and are not crosswalked to reflect border changes, etc.
\begin{itemize}
    \item The five constituent counties of New York City, NY are combined into a New York City, NY entity.
    \item Chicago, IL and the remainder of Cook County, IL, are combined into a complete Cook County, IL.
    \item Kansas City, MO and the rest of Jackson County, MO. are combined into a complete Jackson County, MO.
\end{itemize}


\newpage

\pagenumbering{gobble}
\pagenumbering{arabic}
\renewcommand*{\thepage}{B-\arabic{page}}

\section{LLM Prompt Collection} \label{sec:app_prompts}

Below we present the full, final collection of prompts used to generate digitized and harmonized tables. We settled on the following ordered structure for LLM calls:
\begin{enumerate}
    \item {\bf Main Prompt} is described in the main text. It includes the following variables and subcalls: 
        \begin{itemize}
            \item \verb|{format_instructions}| triggers either {\bf County-Sort Prompt} or {\bf Year-Sort Prompt}
            \item \verb|{additional_instructions}| trigger additional {\bf State-Specific Prompts} 
            \item LLM: Gemini 3.1 Pro with ``structured output'' option selected
        \end{itemize}
    \item {\bf Prompt Columns} and {\bf Prompt Counties} are then used to align the column and county names to a provided set of potential column and county names.
        \begin{itemize}
            \item LLM: Claude Sonnet 4.6
        \end{itemize}
\end{enumerate}

\subsubsection*{County-Sort Prompt}
\begin{quote}

Extraction target

Include a table ONLY if:
\begin{enumerate}
    \item the rows correspond to geographic units such as counties or cities, and
    \item at least one data column reports vehicle registrations or vehicle counts.
\end{enumerate}

Exclude:
\begin{itemize}[label=$-$]
    \item state totals unless they appear as rows inside an otherwise eligible county/city table
    \item tables with no county/city rows
    \item text outside tables
    \item narrative summaries that are not structured tables
\end{itemize}

For each included table, extract:
\begin{itemize}[label=$-$]
    \item \verb|title|: the table title as printed
    \item \verb|column_names|: list of all column headers, INCLUDING the leftmost row-label column
    \item \verb|row_names|: list of row labels, one per data row, EXCLUDING header rows
    \item \verb|data|: rectangular 2D list of strings containing the table body values, EXCLUDING the leftmost row-label column and topmost column-label column
    \item \verb|footnotes|: all footnotes associated with that table, or \verb|``''| if none
\end{itemize}

Required shape rules:

If there are M data rows and N data columns (excluding the row-label column and column-label row), then:
\begin{itemize}[label=$-$]
    \item \verb|len(row_names)| must equal M
    \item \verb|len(column_names)| must equal N + 1
    \item \verb|len(data)| must equal M
    \item every row in data must have length N. You must verify that each row has the same number of cells, each with one value. You must verify that each row has length one less than \verb|column_names|. You must verify that the length of data is the same as the length of \verb|row_names|.
    \item \verb|column_names[0]| must be ``Geography'' whenever the leftmost column contains counties, cities, or other geographic unit names
    \item \verb|row_names| must appear in the same order as in the table
    \item data must appear in the same row order and column order as in the table
\end{itemize}

Header standardization:

For every non-geographic column, format the header exactly as:

\verb|``Field: ENTER_FIELD; Type: ENTER_TYPE; Date: ENTER_DATE; Field\_Standardized: ENTER\_STANDARDIZED\_FIELD''|

For the geographic column, use exactly: Geography

Determine each non-geographic header using all printed header information relevant to that column, including:
\begin{itemize}[label=$-$]
    \item The column label
    \item Any spanner header above it
    \item The table title if needed for context
\end{itemize}

If a column header spans multiple printed header rows, merge them into one final normalized header.

Header field definitions:

Field: the substantive quantity measured by the column, such as automobiles, motor vehicles, trucks, trailers, buses, motorcycles, population, fees, or revenue

Type:
\begin{itemize}[label=$-$]
    \item ``Vehicle'' for vehicle counts or vehicle registration counts
    \item ``Fee'' for monetary registration fees or revenue amounts
    \item ``Other'' otherwise
\end{itemize}

Date must use one of these formats only:
\begin{enumerate}
    \item YYYY
    \item MM/DD/YYYY to MM/DD/YYYY
    \item YYYY (XX months)
    \item Total
\end{enumerate}

Date should be ``Total'' if it refers to the sum over several listed time periods.

\verb|Field\_Standardized|
\begin{itemize}[label=$-$]
    \item Use nan if Type is not ``Vehicle''
    \item otherwise use exactly one of:
\end{itemize}

\begin{enumerate}
    \item Automobiles: motor vehicle primarily designed to transport passengers
    \item Trucks: motor vehicle primarily designed to transport equipment
    \item Buses: large motor vehicle for numerous passengers
    \item Automobiles and Trucks: includes both automobiles and trucks
    \item Automobiles and Buses: includes both automobiles and buses
    \item Trailers: vehicle designed to be towed
    \item Motorcycles
    \item Total Vehicles Registered: Total vehicles. Must include automobiles and trucks, as well as other vehicle categories.
\end{enumerate}

Standardization guidance:

\begin{itemize}[label=$*$]
    \item Multiple columns may map to the same \verb|Field_Standardized| (e.g. if there are multiple types of automobiles or trucks in one table)
    \item Use nan if the vehicle column does not map reliably to one of the allowed standardized values
\end{itemize}

\end{quote}

\subsubsection*{Year-Sort Prompt}
\begin{quote}

Extraction target

Include a table ONLY if:
\begin{enumerate}[label=$-$]
    \item the rows correspond to time units such as years, and
    \item at least one data column reports vehicle registrations or vehicle counts.
\end{enumerate}

Exclude:
\begin{itemize}[label=$-$]
    \item state totals unless they appear as rows/columns inside an otherwise eligible county/city table
    \item tables with no county/city columns
    \item text outside tables
    \item narrative summaries that are not structured tables
\end{itemize}

For each included table, extract:
\begin{itemize}
    \item title: the table title as printed
    \item \verb|column_names|: list of all column headers, INCLUDING the leftmost row-label column
    \item \verb|row_names|: list of row labels, one per data row, EXCLUDING header rows
    \item data: rectangular 2D list of strings containing the table body values, EXCLUDING the leftmost row-label column and topmost column-label column
    \item footnotes: all footnotes associated with that table, or \verb|""| if none
\end{itemize}

Required shape rules:

If there are M data rows and N data columns (excluding the row-label column and column-label row), then:
\begin{itemize}[label=$-$]
    \item \verb|len(row_names)| must equal M
    \item \verb|len(column_names)| must equal N + 1
    \item \verb|len(data)| must equal M
    \item every row in data must have length N. You must verify that each row has the same number of cells, each with one value. You must verify that each row has length one less than \verb|column_names|. You must verify that the length of data is the same as the length of \verb|row_names|.
    \item \verb|column_names[0]| must be \verb|"Geography"| whenever the leftmost column contains counties, cities, or other geographic unit names
    \item \verb|row_names| must appear in the same order as in the table
    \item data must appear in the same row order and column order as in the table
\end{itemize}

Header standardization:

For every non-date column, format the header exactly as:

\verb|Field: ENTER_FIELD; Type: ENTER_TYPE; Geography: ENTER_GEOGRAPHY; Field_Standardized: ENTER_STANDARDIZED_FIELD|

For the date column, use exactly: Date

Determine each non-geographic header using all printed header information relevant to that column, including:
\begin{itemize}[label=$-$]
    \item The column label
    \item Any spanner header above it
    \item The table title if needed for context
\end{itemize}

If a column header spans multiple printed header rows, merge them into one final normalized header.

Header field definitions:

Field: The substantive quantity measured by the column, such as automobiles, motor vehicles, trucks, trailers, buses, motorcycles, population, fees, or revenue. This should be determined based on the column header as well as any relevant information in the table title. Do not include the geography as part of the field.

Type:
\begin{itemize}[label=$-$]
    \item \verb|"Vehicle"|  for vehicle counts, vehicle registration counts, or vehicle-owner counts
    \item \verb|"Fee"| for monetary registration fees or revenue amounts
    \item \verb|"Other"| otherwise
\end{itemize}

Geography: the geographical area to which the column refers to, which could be a county, city or state

\verb|Field_Standardized|
\begin{itemize}[label=$-$]
    \item Use nan if Type is not \verb|"Vehicle"|
    \item otherwise use exactly one of:
\end{itemize}

\begin{enumerate}
    \item Automobiles: motor vehicle primarily designed to transport passengers
    \item Trucks: motor vehicle primarily designed to transport equipment
    \item Buses: large motor vehicle for numerous passengers
    \item Automobiles and Trucks: includes both automobiles and trucks
    \item Automobiles and Buses: includes both automobiles and buses
    \item Trailers: vehicle designed to be towed
    \item Motorcycles
    \item Total Vehicles Registered: Total vehicles. Must include automobiles and trucks, as well as other vehicle categories.
\end{enumerate}

Standardization guidance:

\begin{itemize}[label=$*$]
    \item Multiple columns may map to the same \verb|Field_Standardized| (e.g. if there are multiple types of automobiles or trucks in one table)
    \item Use nan if the vehicle column does not map reliably to one of the allowed standardized values
\end{itemize}

\end{quote}

\subsubsection*{State-Specific Prompts}

{\small 
\begin{tabular}{p{1cm} p{13.7cm}}
MO &	For Missouri, St. Louis City is sometimes listed at the top or bottom of the table as "St. Louis". Make sure to label it as "St. Louis City". Kansas City is also sometimes listed separately and should be labelled as "Kansas City". \\
NY &	The city of New York is often included in the list of counties. Please identify the city of New York as "New York City" and the county of New York (Manhattan) as "New York". \\
IL &	Cook County is often separated into Chicago and the rest of Cook County. If this is the case, label Chicago as "Chicago" and the rest of Cook County as "Cook Excluding Chicago". \\
CA &	Sometimes vehicle registrations are separated into those which paid a fee and those which didn't. Both of these columns should have Type: Vehicle and should be labeled as [Vehicle Type] Fee Paid and [Vehicle Type] Fee Exempt. \\
\end{tabular}
}

\subsubsection*{Prompt Columns}

\begin{quote}
You are a researcher aligning the column names of OCR'd tables with column names processed by humans. You will be given two lists of column names: Assess and True. Please output a Python dictionary where every column in Assess is assigned a corresponding column in True.

Here are some things you should know:
    \begin{itemize} [label=$-$]
        \item   There might be columns in Assess that are not in True and there might be columns in True that are not in Assess. In that case, assign \verb|NoCorrespondingColumnX| to that column, where X is the number of column with that label.
        \item Each column in True may only appear in Assess once.
        \item The matches are not always textually very similar. For example, Automobiles are often recorded as "Cars", "Passenger Cars", "Owners", "Pleasure Cars" etc.
    \end{itemize}
    
You MUST return only a Python dictionary and no other text.

Example Output: \verb|{{"Passenger Cars":"Automobiles"}}|. 

Assess: \verb|{cols_yhat}| 

True: \verb|{cols_y}|.
\end{quote}

\subsubsection*{Prompt Counties}

\begin{quote}

You are a researcher aligning the county names of OCR'd tables with county names processed by humans. You will be given two lists of counties names: Assess and True. Please output a python dictionary where every county in Assess is assigned a corresponding county in True. Don't output any other text.

Here are some things you should know:
    \begin{itemize}[label=$-$]
        \item Every county in True can appear in Assess only once.
        \item Assign None to the county in Assess if it's not in True.
    \end{itemize}

    Assess: \verb|{str_counties_yhat}|
    
    True: \verb|{str_counties_y}|
\end{quote}

\end{document}

%% file: Figures/pipeline/pipeline.tex
\begin{tikzpicture}[
    auto,
    node distance=0.5cm and 0.5cm,
    process/.style={rectangle, draw=black, thick, text width=6cm, minimum height=1cm, align=center, fill=gray!10},
    io/.style={trapezium, draw=black, thick, text width=3.5cm, minimum height=0.8cm, align=center, trapezium left angle=70, trapezium right angle=110, fill=blue!10},
    meta/.style={ellipse, draw=black, thick, text width=3cm, minimum height=0.8cm, align=center, fill=green!10},
    group/.style={draw, dashed, inner sep=4mm, rounded corners},
    line/.style={draw, thick, -{Stealth}, shorten >=3pt}
]

\node[io, text width = 6cm] (input) {Source Table Images};
\node[process, below=of input, text width = 8cm] (preprocess) {\textit{Image Preprocessing}
    \begin{itemize}[noitemsep, topsep=0pt]
        \item Page-rotation detection
    \end{itemize}
};
\node[io, below=of preprocess, text width = 6cm] (crop) {Oriented Table Images};

\node[process, below=3.0cm of crop, text width = 7cm] (llm) {\textit{Multimodal LLM Processing}
    \begin{itemize}[noitemsep, topsep=0pt]
        \item Structured table extraction (headers, rows, cells)
        \item Header Standardization
    \end{itemize}
};
\node[io, below=of llm, text width = 5cm] (raw) {Structured Tables};
\node[process, below=of raw, text width = 7cm] (post) {\textit{Post-processing \& Alignment}
    \begin{itemize}[noitemsep, topsep=0pt]
        \item Residual column and county name standardization
    \end{itemize}
};
\node[io, below=of post] (aligned) {Aligned Tables};

\node[meta, right=0.35cm of preprocess, text width = 3.35cm] (meta0) {OCR and layout-detection models};

\node[meta, right=0.55cm of llm, text width = 6.2cm, yshift=0.25cm] (meta1) {\begin{itemize}[noitemsep, topsep=0pt]
    \item LLM instructions (general, format-specific, and state-specific)
    \item Reference column list (Standardized vehicle categories)
\end{itemize}};

\node[meta, right=0.55cm of post, text width = 6.2cm, yshift=-0.35cm] (meta2) {\begin{itemize}[noitemsep, topsep=0pt]
    \item LLM instructions
    \item Reference column and county lists
\end{itemize}};

\node[group, fit={(input) (preprocess) (crop) (meta0)}] (codeworkflow) {};
\node at ([xshift=-2.5cm, yshift=0cm]codeworkflow.west) {\textbf{Preprocessing Workflow}};

\node[group, fit={(llm) (raw) (post) (aligned) (meta1) (meta2)}] (llmworkflow) {};
\node at ([xshift=-2.5cm, yshift=0cm]llmworkflow.west) {\textbf{LLM-Based Workflow}};

\path[line] (input) -- (preprocess);
\path[line] (preprocess) -- (crop);
\path[line] (crop) -- (llm);
\path[line] (llm) -- (raw);
\path[line] (raw) -- (post);
\path[line] (post) -- (aligned);

\path[line] (meta0.west) -- (preprocess.east);
\path[line] (meta1.west) -- (llm.east);
\path[line] (meta2.west) -- (post.east);
\end{tikzpicture}

%% file: Figures/pipeline/loop.tex
\begin{tikzpicture}[
    auto,
    node distance=0.5cm and 0.8cm,
    process/.style={rectangle, draw=black, thick, text width=6cm, minimum height=1cm, align=center, fill=gray!10},
    io/.style={trapezium, draw=black, thick, text width=3.5cm, minimum height=0.8cm, align=center, trapezium left angle=70, trapezium right angle=110, fill=blue!10},
    meta/.style={ellipse, draw=black, thick, text width=4.5cm, minimum height=1cm, align=center, fill=green!10},
    goldprocess/.style={rectangle, draw=black, thick, text width=6cm, minimum height=1cm, align=center, fill=yellow!30, font=\bfseries},
    goldio/.style={trapezium, draw=black, thick, text width=3.5cm, minimum height=0.8cm, align=center, trapezium left angle=70, trapezium right angle=110, fill=yellow!30, font=\bfseries},
    line/.style={draw, thick, -{Stealth}, shorten >=3pt}
]
\node[process] (llmworkflow) {\textit{LLM-Based Workflow}};
\node[io, below=of llmworkflow, text width = 6cm] (aligned) {Aligned Tables};

\node[process, below=of aligned, text width = 7cm] (validation) {\textit{Validation}
    \begin{itemize}[noitemsep, topsep=0pt]
        \item Cell-level digitization (vs.\ gold standard)
        \item Column harmonization
    \end{itemize}
};
\node[goldio, left=of validation, node distance=6cm] (groundtruth) {Ground Truth};

\node[meta, right=of aligned, node distance=4cm, text width = 6cm] (systemprompts) {
\textbf{Iterative Improvement}
\begin{itemize}[noitemsep, topsep=0pt]
    \item LLM instructions (general, format-specific, and state-specific)
    \item Standardized vehicle categories
    \item Reference column list
    \item Reference county lists
\end{itemize}
};
\path[line] (systemprompts.west) -- (llmworkflow.east);

\path[line] (llmworkflow) -- (aligned);
\path[line] (aligned) -- (validation);
\path[line] (groundtruth) -- (validation);
\path[line] (validation) -- (systemprompts);
\end{tikzpicture}

%% file: Tables/textract_comparison.tex
\begin{tabular}{lr}
\toprule
Metric & Value\\
\midrule
LLM Failure & 0.35\% \\
Textract Failure & 61.40\% \\
\addlinespace
Num.\ of Tables & 285 \\
\bottomrule
\end{tabular}

%% file: Tables/overall_performance_test.tex
\begin{tabular}{lr}
\toprule
Metric & Value\\
\midrule
Total Error Rate (\%) & 4.64\\ 
\quad Missing Output (\%) & 1.36\\ 
\quad Incorrect Output (\%) & 3.29\\ 
\addlinespace
Mean Error (Units) & -7\\
Mean Abs.\ Error (Units) & 15\\
\addlinespace
Mean Percentage Error (\%) & 0.33\\
Mean Abs.\ Percentage Error (\%) & 0.67\\
\addlinespace
$R^2$ (True vs.\ LLM Values) (\%) & 100.00 \\
\addlinespace
Num.\ of Cells & 37,636\\
Num.\ of Tables & 285\\
\bottomrule
\end{tabular}

%% file: Tables/error_only_performance_test.tex
\begin{tabular}{lr}
\toprule
Metric & Value\\
\midrule
Mean Error (Units) & -215\\
Mean Abs.\ Error (Units) & 435\\
\addlinespace
Median Error (Units) & -51\\
Median Abs.\ Error (Units) & 103\\
\addlinespace
Mean Error (\%) & 9.89\\
Mean Abs.\ Percentage Error (\%) & 20.13\\
\addlinespace
Median Percentage Error (\%) & -2.69\\
Median Abs.\ Percentage Error (\%) & 5.52\\
\addlinespace
75th Percentile Abs.\ Error (\%) & 8.42\\
95th Percentile Abs.\ Error (\%) & 25.43\\
\addlinespace
$R^2$ (True vs.\ LLM Values) (\%) & 99.43\\
\addlinespace
Num.\ of Cells & 1,238\\
Num.\ of Tables & 115\\
\bottomrule
\end{tabular}

%% file: Tables/alignment.tex
\begin{tabular}{lcccc}
\toprule
        & Unique  & Unique   & \% of Cells    &  Coincidence \\
        & Headers & Headers  & w/ St.\ Header &  of Fields \\
        & Unst.   & St.      & $=$ `None'     & w/ State Totals \\
        & (1) & (2) & (3) & (4) \\
\midrule
 \multicolumn{4}{l}{\bf A. Coincidence with State Totals} \\
\quad Gold Standard &  76  &  8 & 0.00\% & 88.29\%  \\
\quad LLM Data      &  76  & 13 & 1.23\% & 85.79\% \\
\addlinespace
\hline
\multicolumn{4}{l}{\bf B. Coincidence across Datasets} \\
 \quad Common Unique Headers & \multicolumn{4}{c}{6} \\
 \quad Header Coincidence    & \multicolumn{4}{c}{88.19\%} \\
\bottomrule
\end{tabular}



%% file: Tables/newemp_1.tex
\begin{threeparttable}
\def\sym#1{\ifmmode^{#1}\else\(^{#1}\)\fi}
\begin{tabular}{l*{8}{c}}
\toprule
            & \multicolumn{2}{c}{1920--1930} & \multicolumn{2}{c}{1930--1940} & \multicolumn{2}{c}{1940--1950} & \multicolumn{2}{c}{1950--1960} \\
            \cmidrule(lr){2-3} \cmidrule(lr){4-5} \cmidrule(lr){6-7} \cmidrule(lr){8-9} 
            & $y_{cst}^{\text{LLM}}$ & $y_{cst}$ & $y_{cst}^{\text{LLM}}$ & $y_{cst}$ & $y_{cst}^{\text{LLM}}$ & $y_{cst}$ & $y_{cst}^{\text{LLM}}$ & $y_{cst}$ \\
            &\multicolumn{1}{c}{(1)}&\multicolumn{1}{c}{(2)}&\multicolumn{1}{c}{(3)}&\multicolumn{1}{c}{(4)}&\multicolumn{1}{c}{(5)}&\multicolumn{1}{c}{(6)}&\multicolumn{1}{c}{(7)}&\multicolumn{1}{c}{(8)}\\
\midrule
\multicolumn{8}{l}{{\bf Panel A. Serial Correlation in Adoption}} \\[4pt]
$y_{cs,t-10}^{\text{LLM}}$
            &       0.264**&              &       0.775**&              &       0.722**&              &       0.793**&              \\
            &     (0.063)  &              &     (0.022)  &              &     (0.031)  &              &     (0.022)  &              \\
$y_{cs,t-10}$
            &              &       0.437**&              &       0.772**&              &       0.719**&              &       0.788**\\
            &              &     (0.030)  &              &     (0.023)  &              &     (0.031)  &              &     (0.022)  \\
\grayrule
$p$-value of $H_0: \hat{\rho}^\text{LLM} = \hat{\rho}$       
            & \multicolumn{2}{c}{[0.008]}   & \multicolumn{2}{c}{[0.461]}   & \multicolumn{2}{c}{[0.424]}   & \multicolumn{2}{c}{[0.609]}  \\
$R^2$       &       0.548  &       0.691  &       0.826  &       0.820  &       0.778  &       0.773  &       0.871  &       0.872  \\
N           &         573  &         573  &         543  &         543  &         526  &         526  &         512  &         512  \\
\midrule
\multicolumn{8}{l}{{\bf Panel B. Vehicle Adoption and Population Growth}} \\[4pt]
$\ln(\text{pop}_{cst})$ 
            &       0.281* &       0.286* &       0.090  &       0.087  &      -0.288**&      -0.296**&      -0.205**&      -0.200**\\
            &     (0.117)  &     (0.116)  &     (0.059)  &     (0.060)  &     (0.055)  &     (0.056)  &     (0.027)  &     (0.027)  \\
\grayrule
$p$-value of $H_0: \hat{\beta}^\text{LLM} = \hat{\beta}$    
            & \multicolumn{2}{c}{[0.539]}   & \multicolumn{2}{c}{[0.394]}   & \multicolumn{2}{c}{[0.071]}   & \multicolumn{2}{c}{[0.163]}  \\
$R^2$       &       0.950  &       0.951  &       0.945  &       0.944  &       0.964  &       0.960  &       0.959  &       0.953  \\
N           &        3991  &        3991  &        5783  &        5783  &        3826  &        3826  &        4728  &        4728  \\
\bottomrule
\end{tabular}
\begin{tablenotes}
\item \leavevmode\kern-\scriptspace\kern-\labelsep Panel A of this table presents estimates of persistence in vehicle adoption. We estimate a regression of current log vehicles per capita on the 10- or 11-year lag of log vehicles per capita within one year of each census year. Panel B presents estimates from a regression of current log vehicles per capita on log population separately for each 10-year period. In both panels, the first column for each period presents results using vehicle data from our LLM-based pipeline, and the second column for each period presents results using our gold-standard dataset. For all time periods, we present $p$-values for the test of the null hypothesis, $H_0: \hat{\rho}^{\text{LLM}} = \hat{\rho}$ or $H_0: \hat{\beta}^{\text{LLM}} = \hat{\beta}$, in square brackets. All specifications include state-year fixed effects and those in Panel B include county fixed effects. + $p<0.10$, * $p<0.05$, ** $p<0.01$.
\end{tablenotes}
\end{threeparttable}

%% file: Figures/comparison_with_textract/example_textract.tex
\begin{tabular}{lcccc}
\toprule
COUNTIES. & Passenger Cars. & Commer- cial Cars. & Motor Cycles. & Trailers. \\
\midrule
\rowcolor{gray!20} & & 39 & 3 & 2 \\
Alcona & 733 & 108 & 10 & 4 \\
Alger & 1121 & & 32 & 48 \\
Allegan & 7631 & \cellcolor{gray!20}909 234 & 8 & 27 \\
Alpena & \cellcolor{gray!20}2671 1575 & 95 & 5 & 16 \\
Antrim & & 105 & & 1 \\
Arenac & 1175 & 45 & 1 & \\
Baraga & 697 & 372 & 14 & 31 \\
Barry & 4493 & 1044 & 48 & 32 \\
Bay & \cellcolor{gray!20}9085 1014 & 150 & 1 & 5 \\
Benzie & & & & \\
\rowcolor{gray!20} & & 2308 & 74 & 57 \\
Berricn & 12847 & 424 & 16 & 48 \\
Branch & 5389 & 1366 & 139 & 82 \\
Calhoun & 15483 & 354 & 10 & 12 \\
Cass & 3776 & 242 & 15 & 4 \\
Charlevoix & 2387 & & & \\
\rowcolor{gray!20} & & 158 & 3 & 4 \\
Cheboygan & 1565 & 217 & 9 & 7 \\
Chippewa & 2341 & 82 & 4 & 7 \\
Clare & 934 & 470 & 10 & 82 \\
Clinton & \cellcolor{gray!20}5165 660 & 58 & & 12 \\
\rowcolor{gray!20}Crawford & & & & \\
\bottomrule
\end{tabular}

%% file: Figures/comparison_with_textract/example_llm.tex
\centering
\begin{tabular}{lcccc}
\toprule
COUNTIES & Passenger Cars & Commercial Cars & Motor Cycles & Trailers \\
\midrule
Alcona & 733 & 39 & 3 & 2 \\
Alger & 1121 & 108 & 10 & 4 \\
Allegan & 7631 & 909 & 32 & 48 \\
Alpena & 2671 & 234 & 8 & 27 \\
Antrim & 1575 & 95 & 5 & 16 \\
Arenac & 1175 & 105 & & 1 \\
Baraga & 697 & 45 & 1 & 2 \\
Barry & 4493 & 372 & 14 & 31 \\
Bay & 9085 & 1044 & 48 & 32 \\
Benzie & 1014 & 150 & 1 & 5 \\
Berrien & 12847 & 2308 & 74 & 57 \\
Branch & 5389 & 424 & 16 & 48 \\
Calhoun & 15483 & 1366 & 139 & 82 \\
Cass & 3776 & 354 & 10 & 12 \\
Charlevoix & 2387 & 242 & 15 & 4 \\
Cheboygan & 1565 & 158 & 3 & 4 \\
Chippewa & 2341 & 217 & 9 & 7 \\
Clare & 934 & 82 & 4 & 7 \\
Clinton & 5165 & 470 & 10 & 82 \\
Crawford & 660 & 58 & & 12 \\
\bottomrule
\end{tabular}

%% file: Tables/llm_metrics_fold1-20.tex
\begin{tabular}{lr}
\toprule
Metric & Value\\
\midrule
Total Error Rate (\%) & 2.52\\ 
\quad Missing Output (\%) & 0.91\\ 
\quad Incorrect Output (\%) & 1.61\\ 
\addlinespace
Mean Error (Units) & -6\\
Mean Abs.\ Error (Units) & 10\\
\addlinespace
Mean Percentage Error (\%) & 0.03\\
Mean Abs.\ Percentage Error (\%) & 0.10\\
\addlinespace
$R^2$ (True vs.\ LLM Values) (\%) & 99.99 \\
\addlinespace
Num.\ of Cells & 12,700\\
Num.\ of Tables & 79\\
\bottomrule
\end{tabular}